\documentclass[12pt]{article}
\usepackage{cite,graphicx,epsfig}

\newcommand{\ket}{\left| I \right>}

\newcommand{\state}[1]{\left| \right. \left. #1 \right>}
\newcommand{\ba}{\begin{array}}
\newcommand{\ea}{\end{array}}
\newcommand{\be}{\begin{equation}}
\newcommand{\ee}{\end{equation}}
\newcommand{\bea}{\begin{eqnarray}}
\newcommand{\eea}{\end{eqnarray}}

\def\IB{\relax\hbox{$\inbar\kern-.3em{\rm B}$}}
\def\IC{\relax\hbox{$\inbar\kern-.3em{\rm C}$}}
\def\ID{\relax\hbox{$\inbar\kern-.3em{\rm D}$}}
\def\IE{\relax\hbox{$\inbar\kern-.3em{\rm E}$}}
\def\IF{\relax\hbox{$\inbar\kern-.3em{\rm F}$}}
\def\IG{\relax\hbox{$\inbar\kern-.3em{\rm G}$}}
\def\IGa{\relax\hbox{${\rm I}\kern-.18em\Gamma$}}
\def\IH{\relax{\rm I\kern-.18em H}}
\def\IK{\relax{\rm I\kern-.18em K}}
\def\IL{\relax{\rm I\kern-.18em L}}
\def\IP{\relax{\rm I\kern-.18em P}}
\def\IR{\relax{\rm I\kern-.18em R}}
\def\IZ{\relax{\rm Z\kern-.5em Z}}

\def\half{\frac{1}{2}}

\def\f{\frac}

\def\TL{Temperley-Lieb }

\begin{document}

\begin{titlepage}

\begin{flushright}

\end{flushright}

\vskip 2 cm

\begin{center}
{\LARGE Spectral equivalences and symmetry breaking in integrable $SU_q(N)$ spin chains with boundaries}
\vskip 1 cm

{\large  A. Nichols\footnote{nichols@th.physik.uni-bonn.de} }

\begin{center}
{\it Physikalisches Institut der Universit\"at Bonn, \\
 Nussallee 12, 53115 Bonn, Germany.}
\end{center}

\vskip 1 cm

\vskip .5 cm 

\begin{abstract}
We consider the $SU_q(N)$ invariant spin chain with diagonal and non-diagonal integrable boundary terms.

The algebraic study of spin chains with different types of boundary
terms is used to motivate a set of spectral equivalences between
integrable chains with purely diagonal boundary terms and ones with an
arbitrary non-diagonal term at one end. For each choice of diagonal
boundary terms there is an isospectral one-boundary problem and vice-versa.

The quantum group $SU_q(N)$ symmetry is broken by the presence of a non-diagonal boundary term however one can use the spectral equivalence with the diagonal chain to easily understand the residual symmetries of the system.
\end{abstract}

\end{center}

\end{titlepage}


\newpage
\section{Introduction}
\renewcommand{\theequation}{\arabic{section}.\arabic{equation}}
\setcounter{equation}{0}  
In a previous paper \cite{Nichols:2004fb} we proved a rather surprising spectral equivalence
between an XXZ chain with arbitrary left boundary term and the same XXZ chain
with purely diagonal boundary terms. In this paper we shall show that this
phenomenon extends to all $SU_q(N)$ spin chains  where, due to the many
different choices
of diagonal boundary terms, the structure is much richer.

The XXZ chain with diagonal boundary conditions has been well studied using
the Bethe Ansatz \cite{Alcaraz:1987uk,Alcaraz:1988zr} due to the existence of
a conserved charge $S^z$. For a particular choice of diagonal boundary terms
the chain has an $SU_q(2)$ quantum group symmetry \cite{Pasquier:1989kd}. An alternative understanding of the properties of the quantum group invariant chain comes
from noting that it can be written in terms of the generators of
the Temperley-Lieb (TL) algebra \cite{TemperleyLieb,MartinBook}. This algebra depends on a single parameter. 

The addition of an arbitrary left boundary operator can be understood
algebraically in terms of the
one-boundary Temperley-Lieb algebra (1BTL) \cite{Martin:1992td,Martin:1993jk,MartinWoodcockI,MartinWoodcockII}. Although the
integrable chain now involves three parameters the 1BTL contains only two of
these. The presence of the boundary term breaks the $SU_q(2)$ symmetry however a remnant survives \cite{Delius:2001qh,Delius:2002mv,Nepomechie:2002js,Doikou:2004km}. This element commutes with all elements of the 1BTL
and at generic points of the algebra has the structure and degeneracies of a
$U(1)$ charge whilst for exceptional points it becomes indecomposable \cite{Nichols:2004fb}. 

A consequence of the spectral equivalence between the one-boundary and
diagonal chain is that there must exist a
representation of the 1BTL within the diagonal chain. We would like to stress
that this is extremely surprising as conventionally the diagonal chain is
considered to require \emph{two} boundary terms. By direct calculation at a low
number of sites one can find the explicit form for the generators of the 1BTL
in the diagonal chain (see Appendix C of \cite{Nichols:2004fb}). One finds
that, although both the bulk and boundary generators commute with the diagonal charge
$S^z$, they all become non-local. In a two site example it was realized
in \cite{Nichols:2004fb} that
these non-local expressions could be brought to a form in which the
bulk generators are given by their standard local expressions and the boundary
generator, which still commutes with $S^z$, is non-local. In this paper we
shall describe how to understand, and generalize, this canonical diagonal representation.

For $SU_q(N)$ spin chains, in contrast to the $SU_q(2)$ case, there are many
different choices of integrable diagonal boundary terms\cite{deVega:1992zd}. In this paper we shall use an algebraic approach
\cite{LevyMartin,Doikou:2002ry,AffineDoikou} to construct solutions to the reflection equation. The main
advantage of such an approach, over more direct approaches \cite{Abad:1995ab,Lima-Santos,Yang:2004xi}, is that it
gives a much more transparent way of understanding the structure of possible integrable boundary
terms. Once one completes the `Baxterization' procedure \cite{Jones:1990hq} then any representation
of the algebra can be used to construct a solution to the Yang-Baxter and
reflection equations. We shall show that
each of the possible integrable diagonal boundary terms is related to a different one-boundary
problem and moreover that all diagonal boundary conditions can be related in this way. A
considerable amount of insight into the existence of spectral equivalences can be gained from examining the canonical
diagonal representation in which, as in the $SU_q(2)$ case, all the bulk
generators keep their standard forms but the boundary generator becomes
non-local. This is not quite the representation that give rise to the diagonal chain however it does encode the algebraic properties of the
diagonal representation and allows one to see both the structure of
conserved charges in one-boundary systems and the
possibility of spectral equivalences. This shows that the results in
\cite{Nichols:2004fb} are the first in a large class of equivalences that can
be found between one-boundary and diagonal systems. 

In section \ref{se:1BTLreview} we review the spectral equivalence between the
diagonal and one-boundary \TL chains found in \cite{Nichols:2004fb}. The results from a low
number of sites motivate the discussion of the non-local diagonal representations of
1BTL. There is a canonical diagonal representation in which only the boundary generator is non-local. In order to understand, even in the 1BTL case, this representation it is essential to introduce the braid group structure. In
section \ref{se:Braid} we review the appearance of braid, Hecke, and boundary
algebras from the standpoint of integrability. In most of this paper we shall
be concerned with the braid limits in which spectral parameters
disappear. The full solution, with spectral parameters, will be required in
order to discuss the integrable chains and can be obtained by a process of
Baxterization (see appendix \ref{se:Baxterization}). In section \ref{se:SUN} we discuss the $SU_q(N)$ invariant spin chain
and the possible integrable diagonal and non-diagonal boundary terms. In section
\ref{se:Diagonal} we give an expression for a boundary
generator formed non-locally from the diagonal generators - the canonical diagonal representation. The existence of such
a representation motivates the construction of spectral equivalences between integrable
one-boundary and diagonal chains. In section \ref{se:SUNbreaking} we use these
results to understand how the $SU_q(N)$ symmetry is broken in the presence of
an integrable boundary term. Finally we present our conclusions and some outstanding
questions. In appendix \ref{se:Loop} we give a diagonal representation of 1BTL in
the link pattern basis. This is the representation which was used in \cite{deGier:2003iu}. 
\section{$SU(2)$ spectral equivalences}
\label{se:1BTLreview}
In this section we shall review the results of \cite{Nichols:2004fb}, 
concerning the XXZ model with boundaries, which will be important for this
paper. 

We begin with the integrable $SU_q(2)$ quantum group Hamiltonian \cite{Pasquier:1989kd}: 
\bea \label{eqn:qgHamiltonian}
H^{qg}&=&-\half \left\{ \sum_{i=1}^{L-1} \left( \sigma^x_i \sigma^x_{i+1} + \sigma^y_i \sigma^y_{i+1} - \cos \gamma \sigma^z_i \sigma^z_{i+1} + \cos \gamma \right) + i \sin \gamma \left(\sigma^z_1 - \sigma^z_L \right) \right\}~~~
\eea
The central result, proved in \cite{Nichols:2004fb} using the Bethe ansatz, is
an exact spectral equivalence between the integrable $SU_q(2)$ chain with arbitrary
non-diagonal boundary term at one end:
\bea \label{eqn:Hnd}
H^{nd}&=&\f{\sin \gamma}{\cos \omega +\cos \delta}\left( i \cos \omega \sigma_1^z + \sigma_1^x  - \sin \omega \right)\nonumber\\
&&-\half \left\{ \sum_{i=1}^{L-1} \left( \sigma^x_i \sigma^x_{i+1} +
    \sigma^y_i \sigma^y_{i+1} - \cos \gamma \sigma^z_i \sigma^z_{i+1} + \cos
    \gamma \right) + i \sin \gamma \left(\sigma^z_1 - \sigma^z_L \right)
\right\}~~~
\eea
and the integrable chain with purely diagonal boundary conditions:
\bea \label{eqn:Hd}
H^{d}&=&-\half \left\{ \sum_{i=1}^{L-1} \left( \sigma^x_i \sigma^x_{i+1} + \sigma^y_i \sigma^y_{i+1} - \cos \gamma \sigma^z_i \sigma^z_{i+1} + \cos \gamma \right) \right.\nonumber\\
&&\left.+\sin \gamma \left[\tan \left(\f{\omega+\delta}{2}\right) \sigma_1^z +
    \tan \left(\f{\omega-\delta}{2} \right)\sigma_L^z  +\f{2 \sin \omega}{\cos
      \omega +\cos \delta} \right] \right\}~~~
\eea
For the $SU(2)$ case all choices of boundary terms are in fact
integrable. When we generalize this spectral equivalence to $SU(N)$ with
$N>2$ we shall see that integrability plays a much more restrictive role.

The $SU_q(2)$ quantum group invariant chain (\ref{eqn:qgHamiltonian}) can be written in terms of the Temperley-Lieb algebra:
\bea 
H^{qg}&=&-\sum_{i=1}^{L-1} e_i
\eea
where the generators $e_i$ $(i=1,\cdots,L-1)$ are given by:
\bea \label{eqn:TLgenerators}
e_i= \half \left\{ \sigma^x_i \sigma^x_{i+1} + \sigma^y_i \sigma^y_{i+1} - \cos \gamma \sigma^z_i \sigma^z_{i+1}  + \cos \gamma + i \sin \gamma \left(\sigma^z_i - \sigma^z_{i+1} \right) \right\}
\eea
and obey the relations:
\bea \label{eqn:TL}
e_i e_{i\pm 1} e_i&=&e_i \nonumber \\
e_i e_j& =& e_j e_i \quad |i-j|>1\\    
e_i^2&=&(q+q^{-1})~e_i\nonumber
\eea
%
%
with $q=e^{i \gamma}$. The addition of an integrable boundary term can be expressed in terms of the 1BTL algebra\cite{Martin:1992td,Martin:1993jk,MartinWoodcockI,MartinWoodcockII}. This is an extension of TL involving a boundary operator $e_0$.
\bea \label{eqn:BoundaryTL}
e_1 e_0 e_1&=&e_1 \nonumber \\
e_0^2&=&\f{\sin \omega}{\sin(\omega+\gamma)} e_0 \\
e_0 e_i&=& e_i e_0 \quad i>1 \nonumber
\eea
With the bulk $e_i$ defined in (\ref{eqn:TLgenerators}) one can verify that the expression\cite{Martin:1992td}:
\bea \label{eqn:e0}
e_0&=&-\half \f{1}{\sin(\omega+\gamma)}\left( i \cos \omega \sigma_1^z +  \sigma_1^x  - \sin \omega \right) \nonumber \\
&=&-\half \f{1}{\sin(\omega+\gamma)}\left(
\begin{array}{cc}
i e^{i \omega} & 1  \\
1 & -i e^{-i \omega}
\end{array}
\right) \otimes {\bf 1} \otimes \cdots \otimes {\bf 1}
\eea
obeys the 1BTL (\ref{eqn:BoundaryTL}). The integrable chain is given by:
\bea \label{eqn:TLlb}
H^{nd}&=&-a e_0-\sum_{i=1}^{L-1} e_i
\eea
Now with the parameterization:
\bea \label{eqn:Definitionofa}
a= \f{2 \sin \gamma \sin(\omega+\gamma)}{\cos \omega + \cos \delta}
\eea
we obtain the non-diagonal Hamiltonian (\ref{eqn:Hnd}).

Right boundary terms can also be written using an extension known as the
two-boundary Temperley-Lieb (2BTL) algebra\cite{JanReview,deGier:2005fx}
where, in addition to the 1BTL relations (\ref{eqn:BoundaryTL}), we also have:
\bea \label{eqn:2BoundaryTL}
e_L e_{L-1} e_L&=&e_L \nonumber \\
e_L^2&=&\f{\sin \omega'}{\sin(\omega' +\gamma)} e_L \\
e_L e_i&=& e_i e_L \quad i<L-1 \nonumber
\eea
In contrast to the TL and 1BTL algebras, the 2BTL is infinite dimensional. The
spin chain lives in a particular finite
dimensional quotient \cite{deGier:2003iu,deGier:2005fx}. Here we shall only consider two-boundary cases in which both ends
are diagonal. This is an important limiting case of the above, in which we
take $w=w'=i \infty$, and the left and right diagonal terms are given by:
\bea
e^{d}_0 = \left( \begin{array}{cc} 
q & 0 \\
0 & 0 \\
\end{array} \right) \otimes {\bf 1} \otimes ... {\bf 1} \quad \quad
e^{d}_L = {\bf 1} \otimes  {\bf 1} \otimes ... \otimes \left( \begin{array}{cc} 
0 & 0 \\
0 & q \\
\end{array} \right) 
\eea
Therefore the diagonal Hamiltonian (\ref{eqn:Hd}) can be written as:
\bea \label{eqn:Hd2BTL}
H^{d}&=& -\sum_{i-1}^{L-1} e_i - \sin\gamma \left[ i - \tan\left(\f{\omega-\delta}{2} \right) \right] e^d_0  -\sin\gamma \left[ i + \tan\left(\f{\omega+\delta}{2} \right) \right] e^d_L \quad
\eea
We shall see later that the spectral equivalences in the $SU(N)$ chains involve generalizations of (\ref{eqn:TLlb}) and (\ref{eqn:Hd2BTL}).
\subsection{1BTL diagonal representations}
At generic points one can construct a similarity transformation between the
two Hamiltonians $H^{nd}$ (\ref{eqn:Hnd}) and $H^d$ (\ref{eqn:Hd}). This implies that there
exists a representation of the 1BTL in the diagonal chain. The result at $2$
sites is given by \cite{Nichols:2004fb}:
\bea 
\label{eqn:1BTLDiagonalLocalizedRepn}
e_0&=&\left(
\begin{array}{llll}
\f{\sin \omega}{\sin(\omega+\gamma)} & 0 & 0 & 0 \\ 0 & \f{\sin
\omega}{\sin(\omega+\gamma)} & \f{\sin
\gamma \cos \left(\f{\omega+\delta}{2}\right)}{
\sin(\omega+\gamma) \cos \left(\f{\omega-\delta}{2}\right)} & 0 \\ 0 & 0 & 0 & 0 \\ 0 & 0 & 0 & 0
\end{array}
\right) \quad e_1=\left(
\begin{array}{llll}
0 & 0 & 0 & 0 \\ 0 & \eta & \eta \xi & 0 \\ 0 & 1 & \xi & 0 \\ 0 & 0 &
0 & 0
\end{array}
\right) \eea
where:
\bea \eta&=&\cos \left( \f{\delta-2 \gamma-\omega}{2} \right) \sec
\left( \f{\delta-\omega}{2} \right) \nonumber \\ \xi&=&\cos \left(
\f{\delta+2 \gamma-\omega}{2} \right) \sec \left( \f{\delta-\omega}{2}
\right) \eea
Notice that all parameters \emph{including} $\delta$ appear in both $e_0$ and
$e_1$. We shall call this representation `the \emph{real} diagonal representation' as inserting (\ref{eqn:1BTLDiagonalLocalizedRepn}) into the integrable
Hamiltonian (\ref{eqn:TLlb}) we obtain exactly the diagonal Hamiltonian
(\ref{eqn:Hd}) for the two site case. 

The structure of this representation is difficult to understand due to the
 additional dependence on $\delta$. However one can use the invertible transformation:
\bea \label{eqn:Diagsim} U=\left(
\begin{array}{llll}
1 & 0 & 0 & 0 \\ 0 & 1 & \xi - e^{-i \gamma} & 0\\ 0 & 0 & 1 & 0\\ 0 &
0 & 0 & 1
\end{array}
\right) \eea
to transform the generators:
\bea 
e_i \rightarrow e_i=U e_i U^{-1} \quad \quad  i = 0, 1 
\eea
This preserves $S^z$ and brings the generators into a canonical form:
\bea \label{eqn:diagonal1BTL} e_0&=&\left(
\begin{array}{llll}
\f{\sin \omega}{\sin(\omega+\gamma)} & 0 & 0 & 0 \\ 0 & \f{\sin
\omega}{\sin(\omega+\gamma)} & 1- \f{e^{i \gamma} \sin
\omega}{\sin(\omega+\gamma)} & 0 \\ 0 & 0 & 0 & 0 \\ 0 & 0 & 0 & 0
\end{array}
\right) \quad e_1=\left(
\begin{array}{llll}
0 & 0 & 0 & 0 \\ 0 & e^{i \gamma} & 1 & 0 \\ 0 & 1 & e^{- i \gamma} &
0 \\ 0 & 0 & 0 & 0
\end{array}
\right) \eea
In this form $e_1$ is the standard generator (\ref{eqn:TLgenerators}) and
there is no parameter $\delta$ in either $e_0$ or $e_1$. We shall call this
representation `the \emph{canonical} diagonal representation'. The crucial difference between this representation
(\ref{eqn:diagonal1BTL}) and the `non-diagonal' representation
(\ref{eqn:e0}) of 1BTL is that $e_0$ acts now in the spin chain not
only on the first site but on \emph{all} the sites of the chain. 

The natural framework to understand this canonical diagonal representation is
the braid group. This will allow us to generalize it easily to an arbitrary
number of sites and also to understand similar structures in all $SU_q(N)$
spin chains. In the next section we shall discuss the relevant braid and
boundary algebras that will be required.

In the 1BTL case one can use the loop basis to define another diagonal representation. This is given in  appendix \ref{se:Loop} and is the representation which was used in \cite{deGier:2003iu}.
\section{Braid groups and integrability}
\renewcommand{\theequation}{\arabic{section}.\arabic{equation}}
\setcounter{equation}{0}  
\label{se:Braid}
\subsection{Yang-Baxter equation}
Integrable systems have
an infinite number of conserved charges allowing a large number of properties
to be derived exactly. A sufficient condition for integrability is the
Yang-Baxter (YB) equation\cite{BaxterBook}:
\bea \label{eqn:YB}
R_i(u) R_{i+1}(u+v) R_i(v) &=& R_{i+1}(v) R_i(u+v) R_{i+1}(u) \\
R_i(u) R_j(v) &=& R_j(v) R_i(u) \quad |i-j|>1 \nonumber
\eea
In this paper
we shall be motivated by integrability but will actually mostly work at the
level of the braid limits in which the spectral parameters disappear. In
this case the YB equation becomes the \emph{braid group} generated by elements $g_i$
with $i=1,\cdots, L-1$ obeying:
\bea \label{eqn:braid}
g_i g_{i+1} g_i &=& g_{i+1} g_i g_{i+1} \\
g_i g_j &=& g_j g_i \quad |i-j|>1 \nonumber
\eea
An important simplification occurs when each generator only
has two eigenvalues. These must be the same for each generator and after a rescaling this amounts to
only a single parameter. The resulting algebra, known as the \emph{Hecke
  algebra}, has relations:
\bea \label{eqn:Hecke}
g_i g_{i+1} g_i &=& g_{i+1} g_i g_{i+1} \quad \quad  i=1,\cdots, L-1\\
g_i g_j &=& g_j g_i \quad \quad |i-j|>1 \nonumber \\
(g_i-1)(g_i+q^2)&=&0 \nonumber
\eea
The great simplification of the Hecke algebra, in contrast to the braid group,
comes from its close relation to the symmetric group ($q=1$ point). We shall
set $q=e^{i \gamma}$ as one is mostly interested in cases in which $\gamma$ is
real.

Once one possesses a representation of the Hecke algebra then one can get a
solution to the full Yang-Baxter equation using the process of Baxterization (see appendix \ref{se:FullYB}).

Using the \TL generators (\ref{eqn:TLgenerators}) we can form a representation of the Hecke algebra:
\bea \label{eqn:TLbraidgen}
g_i=1-q e_i
\eea
This is a quotient of the Hecke algebra as these elements also satisfy the identity:
\bea
\label{eqn:TLquotient}
g_i g_{i+1} g_i - g_i g_{i+1} - g_{i+1} g_i + g_i + g_{i+1} - {\bf 1} = 0
\eea 
\subsection{Boundary Yang-Baxter equation}
In general boundary terms do not preserve the conserved charges of the bulk
system and integrability is lost. For the case of a single boundary term at
the L.H.S the sufficient condition for integrability to be maintained is the boundary Yang-Baxter, or reflection, equation\cite{Cherednik:1985vs,Sklyanin:1988yz}:
\bea \label{eqn:fullreflect}
K_0(v) R_1(u+v) K_0(u) R_1(u-v)&=&R_1(u-v) K_0(u) R_1(u+v) K_0(v) \\
R_i(u) K_0(v) &=& K_0(v) R_i(u) \quad |i-j|>1 \nonumber
\eea
Again one can take a
braid limit and, in addition to the braid group relations, we have an extra boundary generator $g_0$ obeying:
\bea \label{eqn:reflect}
g_0 g_1 g_0 g_1 &=& g_1 g_0 g_1 g_0 \\
g_0 g_i &=& g_i g_0 \quad \quad i>1 \nonumber
\eea
This is known as the braid group of type B. If the bulk generators also obey the Hecke
condition (\ref{eqn:Hecke}) then this is called the \emph{affine} Hecke
algebra\cite{RamRamagge}. A further quotient of this system is when $g_0$ also obeys a quadratic relation:
\bea \label{eqn:BtypeHecke}
(g_0-1)(g_0-r^2)&=&0
\eea
The resulting algebra is known as the Hecke algebra of type B. Note the slight difference in conventions between the quadratic relations in (\ref{eqn:Hecke}) and
(\ref{eqn:BtypeHecke}). For this quotient a process of Baxterization can
again be carried out (see appendix \ref{se:FullHeckeBYB}) to get a solution to
the full reflection equation \cite{LevyMartin,Doikou:2002ry}.

The 1BTL algebra (\ref{eqn:BoundaryTL}) is a representation of the Hecke algebra of type B (with $r=e^{-i \omega}$). In addition to the \TL braid generators (\ref{eqn:TLbraidgen}) we take:
\bea
g_0={\bf 1} -2ie^{-i \omega}\sin(\gamma+\omega) e_0
\eea
This is a quotient of the Hecke algebra of type B as it also obeys the identity:
\bea
g_1 g_0 g_1 - g_1 g_0 -g_0 g_1 + (1+e^{-2i \omega}) g_1 +g_0 - (1+e^{-2i \omega}) {\bf 1} = 0
\eea

The solutions to the reflection equation that will appear in this paper require going beyond the quadratic quotient (\ref{eqn:BtypeHecke}) to a cubic one:
\bea \label{eqn:cubic}
g_0(g_0-1)(g_0-r^2)&=&0 
\eea
with the additional relation:
\bea \label{eqn:extraquotient}
g_1 g_0^2 g_1 g_0^2 -g_0^2 g_1 g_0^2 g_1&=&(1+r^2)(g_1 g_0^2 g_1 g_0 -g_0 g_1 g_0^2 g_1)
\eea
We are not aware of previous discussion of this quotient. The Baxterization
and related integrable model is discussed in appendix
\ref{se:FullNonHeckeBYB}. Although the Hecke algebra of type B is contained
within this quotient we have chosen to present it separately as it is an important subset in which many relations simplify dramatically.

It is also possible to add a boundary term to the right end of the chain. In
this case the braid limit of the reflection equation is:
\bea
g_L g_{L-1} g_L g_{L-1} &=& g_{L-1} g_L g_{L-1} g_L \\
g_L g_i &=& g_i g_L \quad \quad i<L-1 \nonumber
\eea
With boundary generators at both ends we have the affine braid group of type
B.

In a physical system it is common that the generators are realized
locally. For example in a standard spin chain bulk terms $g_i$ are nearest neighbour
interactions and the boundary operators $g_0$ (and $g_L$) act only on the
first (and last) site. However in the abstract formalism of the braid group such a local realization is certainly \emph{not} a requirement.
\section{Integrable $SU_q(N)$ spin chains with boundaries}
\label{se:SUN}
In this section we shall discuss the $SU_q(N)$ spin chains and possible integrable boundary terms that can be added to them.
\subsection{Bulk $SU_q(N)$ invariant chain}
The bulk generators \cite{Jimbo:1989qm} are given in terms of a local interaction on sites $i$ and $i+1$:
\bea
g_i={\bf 1} \cdots \otimes {\bf 1} \otimes A \otimes {\bf 1} \cdots \otimes
{\bf 1} \quad \quad 1 \le i \le L-1
\eea
where:
\bea \label{eqn:SUNRmatrix}
A= \sum_{n=1}^N e_{n,n} \otimes e_{n,n} +q \sum_{n \ne m} e_{n,m} \otimes e_{m,n} + (1-q^2) \sum_{n <m} e_{m,m} e_{n,n}
\eea
and the $e_{m,n}$ are the elementary matrices being only non-zero in the entry
of the $m$'th row and $n$'th column.  It is simple to verify that the $g_i$ obeys the Hecke algebra conditions (\ref{eqn:Hecke}). These generators are invariant under the $SU_q(N)$ quantum group symmetry (see section \ref{se:SUNbreaking}). 

The corresponding integrable chain is found by Baxterizing this braid matrix
to get the $R$-matrix, then forming the full transfer matrix, and finally
extracting the integrable Hamiltonian from this\footnote{When referring to \emph{the}
  integrable Hamiltonian we shall always mean the simplest one.}. The result
(see appendix \ref{eqn:FullYBsoln}) is given by:
\bea \label{eqn:BulkIntegrable}
H=\sum_{i=1}^{L-1} g_i
\eea
There are several different types of integrable boundary terms that can be added to this
Hamiltonian. In this paper we shall consider two general classes: diagonal boundary terms added to both ends and a non-diagonal boundary term added to just one end.
\subsection{Diagonal boundary terms}
\label{se:Diagonalboundaries}
We first consider diagonal boundary terms added to the $SU_q(N)$ integrable chain. Here we shall focus on the braid limit as the full solution can be reconstructed by Baxterizing these solutions. The integrable chain with diagonal boundary terms will be given at the end of this subsection.

We consider an arbitrary diagonal matrix $K^{d}_0$ acting only on the first site:
\bea
g_0=K^{d}_0 \otimes {\bf 1} \cdots \otimes {\bf 1}
\eea
The general diagonal solution to the reflection equation for $SU_q(N)$ chains was given in \cite{deVega:1992zd}. The braid limit of these solutions gives the matrix $g_0$. All the solutions found for a left boundary $g_0^d$ satisfy (up to rescaling):
\bea
g_0^d \left( g_0^d -1 \right) =0
\eea
Therefore $g_0^d$ can only have eigenvalues $0$, and $1$. The different
possibilities are distinguished by $Tr(g_0)$ or equivalently the multiplicity
of each eigenvalue. We shall denote by $g_0^{d~(k)}$ with $k=1,\cdots, N-1$
the solution with $k$ eigenvalues equal to $1$. The general formula for
$K_0^{d~(k)}$ is:
\bea
K_0^{d~(k)} =\sum_{n=1}^k e_{n,n}
\eea
We can also add an integrable diagonal term to the right end:
\bea
g_L= {\bf 1} \cdots \otimes {\bf 1} \otimes K^{d}_L
\eea
Again the
generators $g_L$ only have two eigenvalues and the diagonal cases can have (up
to rescaling) only eigenvalues $0$ and $1$. We denote the different solutions
by  $g_L^{d~(k)}$ with $k=1,\cdots, N-1$ the solution with $k$ eigenvalues equal to $1$. The general formula is given by:
\bea
K_L^{d~(k)} =\sum_{n=N-k+1}^N e_{n,n}
\eea
The first few cases of the left and right diagonal boundary terms are given by:
\subsubsection{$SU(2)$}
In this case there is only one solution:
\begin{itemize}
\item{$k=1$}
\bea
K_0^{d~(1)}=\left( \begin{array}{cc}
1 & 0 \\
0 & 0
\end{array}
\right)  
\quad \quad
K_L^{d~(1)}=\left( \begin{array}{cc}
0 & 0 \\
0 & 1
\end{array}
\right)  \quad {\rm Eigenvalues:}\quad 0,1
\eea
\end{itemize}
\subsubsection{$SU(3)$}
There are now two solutions:
\begin{itemize}
\item{$k=1$}
\bea 
K_0^{d~(1)}=\left( \begin{array}{ccc}
1 & 0 & 0 \\
0 & 0 & 0 \\
0 & 0 & 0 
\end{array} 
\right) 
\quad \quad
K_L^{d~(1)}=\left( \begin{array}{ccc}
0 & 0 & 0 \\
0 & 0 & 0 \\
0 & 0 & 1 
\end{array} 
\right) 
\quad {\rm Eigenvalues:} \quad 0,0,1
\eea
\item{$k=2$}
\bea 
K_0^{d~(2)}=\left( \begin{array}{ccc}
1 & 0 & 0 \\
0 & 1 & 0 \\
0 & 0 & 0 
\end{array} 
\right)
\quad \quad 
K_L^{d~(2)}=\left( \begin{array}{ccc}
0 & 0 & 0 \\
0 & 1 & 0 \\
0 & 0 & 1 
\end{array} 
\right)  \quad {\rm Eigenvalues:} \quad 0,1,1
\eea
\end{itemize}
\subsubsection{$SU(4)$}
There are now three solutions:
\begin{itemize}
\item{$k=1$}
\bea 
K_0^{d~(1)}=\left( \begin{array}{cccc}
1 & 0 & 0 & 0\\
0 & 0 & 0 & 0\\
0 & 0 & 0 & 0 \\
0 & 0 & 0 & 0
\end{array} 
\right)
\quad 
K_L^{d~(1)}=\left( \begin{array}{cccc}
0 & 0 & 0 & 0\\
0 & 0 & 0 & 0\\
0 & 0 & 0 & 0 \\
0 & 0 & 0 & 1
\end{array} 
\right)
~{\rm Eigenvalues:~} 0,0,0,1 \nonumber
\eea
\item{$k=2$}
\bea 
K_0^{d~(2)}=\left( \begin{array}{cccc}
1 & 0 & 0 & 0\\
0 & 1 & 0 & 0\\
0 & 0 & 0 & 0 \\
0 & 0 & 0 & 0
\end{array} 
\right)
\quad 
K_L^{d~(2)}=\left( \begin{array}{cccc}
0 & 0 & 0 & 0\\
0 & 0 & 0 & 0\\
0 & 0 & 1 & 0 \\
0 & 0 & 0 & 1
\end{array} 
\right)~{\rm Eigenvalues:~} 0,0,1,1 \nonumber
\eea
\item{$k=3$}
\bea 
K_0^{d~(3)}=\left( \begin{array}{cccc}
1 & 0 & 0 & 0\\
0 & 1 & 0 & 0\\
0 & 0 & 1 & 0 \\
0 & 0 & 0 & 0
\end{array} 
\right)
\quad
K_L^{d~(3)}=\left( \begin{array}{cccc}
0 & 0 & 0 & 0\\
0 & 1 & 0 & 0\\
0 & 0 & 1 & 0 \\
0 & 0 & 0 & 1
\end{array} 
\right)
~{\rm Eigenvalues:~} 0,1,1,1 \nonumber
\eea
\end{itemize}
The integrable chain with diagonal boundary terms on both ends is parameterized by
integers $(k_1,k_L)$ where $(1 \le k_1,k_L \le N-1)$. It is given by:
\bea \label{eqn:DiagonalIntegrable}
H=a_0 g_0^{d~(k_1)} + \sum_{i=1}^{L-1} g_i + a_L g_L^{d~(k_L)}
\eea
where $a_0$ and $a_L$ are two arbitrary parameters and the solutions $g_0^{d~(k)}$ and $g_L^{d~(k)}$ are those given above. 
\subsection{Non-diagonal Hecke type boundary}
\label{se:NonDiagonalHecke}
For the non-diagonal cases we shall only examine the case of a single left
boundary generator. There are many solutions to the reflection equation
related by global gauge transformations and here we shall give the
simplest. Indeed one of the main benefits of following an algebraic approach
is that it is only sensitive to the real physical parameters of the problem. The addition of a second boundary would restrict the number of
possible gauge transformations forcing one to include more physical parameters.

We found that there are two different types of non-diagonal boundary
generators which are distinguished by the number of
eigenvalues. In this section will shall give the
Hecke type ones in which we have have only two eigenvalues. In the next
section the non-Hecke ones with three different eigenvalues will be given. We did not find any other solutions.

The solutions lying in the type B Hecke
quotient (\ref{eqn:BtypeHecke}) have only two eigenvalues: $1$ and $r^2$. The different
possibilities are distinguished by $Tr(g_0)$ or equivalently the multiplicity
of each eigenvalue. We shall denote by $g_0^{(k)}$ with $k=1,\cdots, N-1$
the solution with $k$ eigenvalues equal to $1$. The boundary generator is given by:
\bea
g_0^{(k)}=K_0^{(k)} \otimes {\bf 1} \otimes {\bf 1} \cdots \otimes {\bf 1}
\eea
The explicit form is given by:
\bea \label{eqn:Heckeboundary}
K^{(k)}_{0}=\sum_{n=1}^{k} e_{n,n} + r^2 \sum_{n=1}^{N-k} e_{n,n} - ir \sum_{n=1}^{min(k,N-k)} \left( e_{n,N+1-n} + e_{N+1-n,n} \right) 
\eea
These were found by directly solving equation (\ref{eqn:reflect}). The general form of these is not so revealing so let us look at the first few cases:
\subsubsection{$SU(2)$}
There is only one solution with $k=1$:
\bea \label{eqn:SU2Heckeboundary}
K_0^{(1)}=\left( \begin{array}{cc}
1+r^2 & -ir \\
-ir & 0
\end{array}
\right) \quad {\rm Eigenvalues:} \quad 1,r^2
\eea
\subsubsection{$SU(3)$}
There are now two solutions:
\begin{itemize}
\item{$k=1$}
\bea \label{eqn:SU31rr} 
K_0^{(1)}=\left( \begin{array}{ccc}
1+r^2 & 0 & -ir \\
0 & r^2 & 0 \\
-ir & 0 & 0 
\end{array} 
\right) \quad {\rm Eigenvalues:} \quad 1,r^2,r^2
\eea
\item{$k=2$}
\bea 
K_0^{(2)}=\left( \begin{array}{ccc}
1+r^2 & 0 & -ir \\
0 & 1 & 0 \\
-ir & 0 & 0 
\end{array} 
\right) \quad {\rm Eigenvalues:} \quad 1,1,r^2
\eea
\end{itemize}
\subsubsection{$SU(4)$}
There are now three solutions:
\begin{itemize}
\item{$k=1$}
\bea 
K_0^{(1)}=\left( \begin{array}{cccc}
1+r^2 & 0 & 0 & -ir \\
0 & r^2 & 0 & 0 \\
0 & 0 & r^2 & 0 \\
-ir & 0 & 0 & 0 
\end{array} 
\right) \quad {\rm Eigenvalues:} \quad 1,r^2,r^2,r^2
\eea
\item{$k=2$}
\bea 
K_0^{(2)}=\left( \begin{array}{cccc}
1+r^2 & 0 & 0 & -ir \\
0 & 1+r^2 & -ir & 0 \\
0 & -ir & 0 & 0 \\
-ir & 0 & 0 & 0 
\end{array} 
\right) \quad {\rm Eigenvalues:} \quad 1,1,r^2,r^2
\eea
\item{$k=3$}
\bea 
K_0^{(3)}=\left( \begin{array}{cccc}
1+r^2 & 0 & 0 & -ir \\
0 & 1 & 0 & 0 \\
0 & 0 & 1 & 0 \\
-ir & 0 & 0 & 0 
\end{array} 
\right) \quad {\rm Eigenvalues:} \quad 1,1,1,r^2
\eea
\end{itemize}
Note that the previous diagonal solutions in section \ref{se:Diagonalboundaries} are obtained from these by
setting $r=0$.

In this case the most general integrable chain (see appendix
\ref{se:FullHeckeBYB}) with $(1 \le k < N)$ can be written as:
\bea
H=-i \f{e^{i(\gamma+\omega)}\sin \gamma}{\cos \omega + \cos \delta} g^{(k)}_0 + \sum_{i=1}^{L-1} g_i 
\eea
where $\delta$ is arbitrary and we take $q=e^{i \gamma}$ and $r=e^{-i
  \omega}$. The parameterization is of course arbitrary and we have chosen one
for later convenience. Note that as $w \rightarrow -i \infty$ (i.e. $r
  \rightarrow 0$) we recover
the integrable chain with an arbitrary left diagonal boundary i.e. (\ref{eqn:DiagonalIntegrable}) with $a_L=0$.
\subsection{Non-Hecke type boundary}
\label{se:NonDiagonalNonHeckeboundaries}
We also found solutions to the reflection equation which did not lie within the Hecke quotient \ref{eqn:Hecke}. These all lie
within a cubic quotient:
\bea
g_0(g_0-1)(g_0-r^2)=0
\eea
Therefore $g_0$ can only have eigenvalues $0$, $1$, and $r^2$. The different
possibilities are distinguished by $Tr(g_0)$ and $Tr(g^2_0)$ or equivalently
the multiplicity of each eigenvalue. We shall denote by $g_0^{(k_1,k_2)}$ the
solution with $k_1$ eigenvalues equal to $1$ and $k_2$ eigenvalues equal to
$r^2$. Clearly we must have $k_1+k_2 < N$ otherwise we have no zero eigenvalues. We shall see shortly that our solutions also obey an
additional relation (\ref{eqn:additionalquot}). The Baxterization of these solutions is given in appendix \ref{se:FullNonHeckeBYB}. For this paper the only purpose of this process is to find the corresponding integrable Hamiltonian.
The boundary generator is given by:
\bea
g_0^{(k_1,k_2)}=K_0^{(k_1,k_2)} \otimes {\bf 1} \otimes {\bf 1} \cdots \otimes {\bf 1}
\eea
The explicit form is given by:
\bea \label{eqn:NonHeckeboundary}
K^{(k_1,k_2)}_{0}=\sum_{n=1}^{k_1} e_{n,n} + r^2 \sum_{n=1}^{k_2} e_{n,n} - ir \sum_{n=1}^{min(k_1,k_2)} \left( e_{n,k_1+k_2+1-n} + e_{k_1+k_2+1-n,n} \right) 
\eea
These were again found by directly solving equation (\ref{eqn:reflect}). Note that taking $k_1+k_2=N$, the case in which $g_0$ has no zero eigenvalues, we reproduce the previous Hecke boundary expression (\ref{eqn:Heckeboundary}).

Let us again examine the first few cases:
\subsubsection{$SU(2)$}
All solutions for $SU(2)$ are of Hecke type (\ref{eqn:SU2Heckeboundary}).
\subsubsection{$SU(3)$}
We have a single solution given by:
\bea \label{eqn:NonHeckeSU311}
K_0^{(1,1)}=\left( \begin{array}{ccc}
1+r^2 & -ir  & 0 \\
-ir & 0 & 0 \\
0 & 0 & 0 
\end{array} 
\right) \quad {\rm Eigenvalues:} \quad 0,1,r^2
\eea
\subsubsection{$SU(4)$}
\begin{itemize}
\item{$k_1=1, k_2=1$}
\bea
K_0^{(1,1)}=\left( \begin{array}{cccc}
1+r^2 & -ir & 0 & 0 \\
-ir & 0 & 0 & 0 \\
0 & 0 & 0 & 0 \\
0 & 0 & 0 & 0 
\end{array} 
\right) \quad {\rm Eigenvalues:} \quad 0,0,1,r^2
\eea
\item{$k_1=1, k_2=2$}
\bea 
K_0^{(1,2)}=\left( \begin{array}{cccc}
1+r^2 & 0 & -ir & 0 \\
0 & r^2 & 0 & 0 \\
-ir & 0 & 0 & 0 \\
0 & 0 & 0 & 0 
\end{array} 
\right) \quad {\rm Eigenvalues:} \quad 0,1,r^2,r^2
\eea
\item{$k_1=2, k_2=1$}
\bea 
K_0^{(2,1)}=\left( \begin{array}{cccc}
1+r^2 & 0 & -ir & 0 \\
0 & 1 & 0 & 0 \\
-ir & 0 & 0 & 0 \\
0 & 0 & 0 & 0 
\end{array} 
\right) \quad {\rm Eigenvalues:} \quad 0,1,1,r^2
\eea
\end{itemize}
Note that these solutions can be obtained by embedding the previous Hecke
solutions into a larger matrix. For example in the case  $k_1=1, k_2=2$ we
have embedded the matrix (\ref{eqn:SU31rr}) into a $4 \times 4$ matrix. This
pattern persists and in this way one obtains all the non-Hecke solutions from
the previous Hecke ones. This observation implies that the matrices satisfy
additional relations. The combination:
\bea \label{eqn:DefinitionofX}
X=(1+r^{2})g_0-g^2_0
\eea
is just a constant matrix $r^2 {\bf 1} $ for the Hecke type solutions. For the
non-Hecke solutions X is equal to $r^2 g^{d~(k_1+k_2)}_0$ where the
$g^{d~(k)}_0$ is as defined in section \ref{se:Diagonalboundaries}. Therefore
as $g^{d~(k)}_0$ satisfies the reflection equation (\ref{eqn:reflect}) we must have the relation:
\bea
X g_1 X g_1 &=& g_1 X g_1 X 
\eea
Inserting the form of $X$ (\ref{eqn:DefinitionofX}) and using the reflection equation (\ref{eqn:reflect}) we find that:
\bea \label{eqn:additionalquot}
g_1 g^2_0 g_1 g^2_0 -(1+r^2) g_1 g^2_0 g_1 g_0= g^2_0 g_1 g^2_0 g_1 -(1+r^2) g_0 g_1 g^2_0 g_1
\eea
This additional relation is necessary to Baxterize the solutions. The most general integrable chain $(1 \le k_1+k_2 < N)$ can be written as (see appendix \ref{se:FullNonHeckeBYB}):
\bea
H^{nd}&=&-i \f{e^{-i(\delta-\gamma)} \sin \gamma }{\cos \omega + \cos \delta} \left( 1+e^{i(\omega+\delta)} +e^{2i \omega} \right) g^{(k_1,k_2)}_0  \\
&&\quad +i \f{e^{-i(\delta-\gamma)} e^{2i \omega} \sin \gamma }{\cos \omega + \cos \delta}  (g^{(k_1,k_2)}_0)^2  + \sum_{i-1}^{L-1} g_i
\eea
where $\delta$ is arbitrary and, as before, we take $q=e^{i \gamma}$ and $r=e^{-i \omega}$.
\section{Diagonal representations and the occurrence of spectral equivalences}
\label{se:Diagonal}
In this section we shall examine the braid and Hecke algebras with one and two-boundary
extensions. We shall show than in certain cases the two boundary algebra contains a one parameter family
of one-boundary algebras. This algebraic
phenomenon will motivate the construction of a set of spectral equivalences between arbitrary integrable one-boundary and diagonal chains. 
\subsection{Diagonal representations}
Let us begin with the braid group generators (\ref{eqn:braid}) supplemented by a
boundary operator $G_0$ at the left end and $G_L$ at the right end. We have:
\bea
g_1 G_0 g_1 G_0 = G_0 g_1 G_0 g_1 \quad \quad g_{L-1} G_{L} g_{L-1} G_{L} =G_{L} g_{L-1} G_{L} g_{L-1} 
\eea
We can `braid translate' the operator $G_{L}$ to the left hand end to
produce a non-local generator:
\bea
G_0^{(NL)} = g_1^{-1} g_2^{-1} \cdots g_{L-1}^{-1} G_L g_{L-1} \cdots g_2 g_1
\eea
It is simple to verify using only the braid relations that this obeys the reflection equation:
\bea
g_1 G_0^{(NL)} g_1 G_0^{(NL)}  = G_0^{(NL)}  g_1 G_0^{(NL)}  g_1
\eea
and commutes with the rest of the bulk generators:
\bea
\left[g_i, G_0^{(NL)} \right] &=&0 \quad \quad 2 \le i \le L-1 
\eea
Now from these two simple solutions $G_0$ and $G_0^{(NL)}$ we can try to form
a one parameter family of solutions to the reflection equation:
\bea \label{eqn:NonLocalFamily}
g_0=G_0 + r^2G_0^{(NL)}
\eea
This will obey the reflection equation if:
\bea \label{eqn:LocalNonLocalRE}
g_1 G_0 g_1 G_0^{(NL)} + g_1 G_0^{(NL)}  g_1 G_0 = G_0 g_1
G_0^{(NL)} g_1 + G_0^{(NL)} g_1 G_0 g_1
\eea
Now it is useful to rewrite $G_0^{(NL)}=g_1^{-1} Y g_1$ where $Y$ commutes
with $G_0$. Then we have:
\bea 
g_1^2 G_0 Y g_1  + g_1 Y g_1^2 G_0 = g_1 G_0 Y g_1^2 +  Y g_1^2 G_0 g_1
\eea
Up to this point we have used purely braid group relations. However if $g_1$ also
obeys the Hecke condition (\ref{eqn:Hecke}) then this simplifies to:
\bea
g_1 Y g_1 G_0 = Y g_1 G_0 g_1
\eea
which is equivalent to:
\bea
\left[ g_1, G_0^{(NL)} G_0 \right] =0
\eea
We shall examine the case in which a stronger condition is obeyed:
\bea 
G_0^{(NL)} G_0 =0
\eea
As the bulk generators are invertible the condition
(\ref{eqn:OnebraidQuotient}) is equivalent to:
\bea \label{eqn:OnebraidQuotient}
G_L g_{L-1} \cdots g_2 g_1 G_0 =0
\eea
Therefore if we have integrable boundary terms on the left and right ends
obeying (\ref{eqn:OnebraidQuotient}) then we can construct a family of
non-local solutions to the reflection equation (\ref{eqn:reflect}). We have
assumed throughout that the bulk braid group generators $g_i$ are
invertible. However the boundary generators $G_0$ and $G_L$ cannot be invertible
elements otherwise the condition (\ref{eqn:OnebraidQuotient}) reduces to a trivial one.

Let us specialize to the case of interest for this paper in which
both the boundary generators are diagonal. In this case they satisfy:
\bea
G_0(G_0-1)=0 \quad \quad  G_L(G_L-1)=0  
\eea
and therefore we have:
\bea
G_0^{(NL)}(G_0^{(NL)}-1)=0
\eea
Let us now form the powers of the generator $g_0$ (\ref{eqn:NonLocalFamily}):
\bea
g_0^2&=& G_0 + r^2 G_0  G_0^{(NL)} + r^4 G_0^{(NL)} \\
g_0^3&=& G_0 + (r^2+r^4) G_0 G_0^{(NL)} + r^6G_0^{(NL)}
\eea
We see that we must have the identity:
\bea
g_0 (g_0-1)(g_0-r^2) = 0
\eea
Now we can form the combination $X$ as in (\ref{eqn:DefinitionofX}):
\bea
X=(1+r^{2})g_0-g^2_0=r^2\left(G_0+G_0^{(NL)}-G_0 G_0^{(NL)} \right)
\eea
By using the Hecke condition for $g_1$ (\ref{eqn:Hecke}) and the condition
(\ref{eqn:OnebraidQuotient}) one can verify, after a great deal of simple but
tedious algebra, that algebraically this quantity $X$ also obeys the reflection
equation. We did not find a simple way to demonstrate this fact. Therefore the
generator $g_0$ given by (\ref{eqn:NonLocalFamily}) also obeys the additional relation (\ref{eqn:additionalquot}).

In the cases in which we have the more restrictive condition:
\bea
(g_0-1)(g_0-r^2) = 0
\eea
we have:
\bea \label{eqn:MorerestrictiveOnebraidQuotient}
(G_0-1)(G_0^{(NL)}-1)=0
\eea
We have shown that from the two-boundary
braid algebra in which the bulk generators obey the Hecke condition and the
boundary ones obey (\ref{eqn:OnebraidQuotient}) we can form a one parameter
set of non-local solutions to the reflection equation. Conversely if we are given a
one-parameter family of solutions, parameterized by $r$, from (\ref{eqn:NonLocalFamily}) we can simply
read off the individual generators $G_0$ and $G_0^{(NL)}$. Algebraically these
non-local solutions behave
in exactly the same way as the non-diagonal solutions of sections
\ref{se:NonDiagonalHecke} and \ref{se:NonDiagonalNonHeckeboundaries}.

It remains for us to discuss \emph{when} the integrable diagonal boundary
terms of section \ref{se:Diagonalboundaries} obey the constraint
(\ref{eqn:OnebraidQuotient}). This is the case when $k_1 + k_2 \le N$. All of
the one-boundary non-diagonal terms given in sections
\ref{se:NonDiagonalHecke} and \ref{se:NonDiagonalNonHeckeboundaries} obey this
condition. The more restrictive constraint
(\ref{eqn:MorerestrictiveOnebraidQuotient}) is obeyed when $k_1 + k_2 = N$
which is exactly the point at which the non-Hecke boundary terms reduce to the
purely Hecke ones. We shall show in the next subsection that in all the cases
in which $k_1 + k_2 \le N$ there is an exact spectral equivalence between the the diagonal and one-boundary Hamiltonians. We shall return to the question of the diagonal terms which have $k_1 + k_2 > N$ shortly.
\subsection{Spectral equivalences}
\label{sec:SpectralEquivs}
The previous algebraic considerations have shown us that for every set of
diagonal generators obeying the condition (\ref{eqn:OnebraidQuotient}) one can
form a non-local family of solutions to the reflection equation. Such
solutions are algebraic as they depend only on the parameters entering the
algebra and obey the same relations as the general non-diagonal solutions. This is the generalization of the canonical diagonal representation
(\ref{eqn:diagonal1BTL}) from the XXZ case. As explained in section \ref{se:1BTLreview} this is distinct from the real diagonal representation (\ref{eqn:1BTLDiagonalLocalizedRepn}) that also depends on the parameters in the Hamiltonian.

Although we have not managed to understand the structure of the real diagonal representation we were motivated by the existence of the canonical diagonal representation to search for spectral equivalences between one-boundary and diagonal Hamiltonians. These have been verified numerically and by explicit diagonalization at a low number of sites.
\subsubsection{Hecke type boundary}
There is a spectral equivalence between the following $SU(N)$ integrable chains:
\bea
\label{eqn:Heckespectralequiv}
H^{Hecke,nd}=-i \f{e^{i(\gamma+\omega)}\sin \gamma}{\cos \omega + \cos \delta} g^{(k)}_0 -\f{i e^{i(\gamma-\delta)}\sin \gamma}{\cos \omega + \cos \delta} + \sum_{i-1}^{L-1} g_i
\eea
and:
\bea
H^{Hecke,d}&=& \sum_{i-1}^{L-1} g_i - e^{i \gamma}\sin\gamma \left[ i - \tan\left(\f{\omega-\delta}{2} \right) \right] g^{d ~(  k)}_0 \nonumber \\
&&\quad -e^{i \gamma}\sin\gamma \left[ i + \tan\left(\f{\omega+\delta}{2} \right) \right] g^{d ~(N-k)}_L 
\eea
where the non-diagonal Hecke boundary term $g^{(k)}_0$ was given in section \ref{se:NonDiagonalHecke} and the diagonal terms $g^{d ~(  k)}_0$ and $g^{d ~(N-k)}_L$ were given in \ref{se:Diagonalboundaries}.

For the $SU(2)$ case the braid operators can be expressed in terms of the 1BTL generators:
\bea
g_i&=& {\bf 1} -e^{i \gamma} e_i\\
g^{(1)}_0&=&{\bf 1} -2ie^{-i \omega}\sin(\gamma+\omega) e_0\\
g_0^d&=&\f{1}{2} \left(1+\sigma_1^z \right) \\
g_L^d&=&\f{1}{2} \left(1-\sigma_L^z \right)
\eea
Inserting these definitions into the above equations we reproduce the spectral
equivalence found in the XXZ model - see section \ref{se:1BTLreview}. 
\subsubsection{Non-Hecke type boundary}
There is a spectral equivalence between the following $SU(N)$ integrable chains:
\bea
\label{eqn:NonHeckespectralequiv}
H^{Non-Hecke,nd}&=&-i \f{e^{-i(\delta-\gamma)} \sin \gamma }{\cos \omega + \cos \delta} \left( 1+e^{i(\omega+\delta)} +e^{2i \omega} \right) g^{(k_1,k_2)}_0  \nonumber \\
&& \quad +i \f{e^{-i(\delta-\gamma)} e^{2i \omega} \sin \gamma }{\cos \omega + \cos \delta}  (g^{(k_1,k_2)}_0)^2  + \sum_{i-1}^{L-1} g_i
\eea
and:
\bea
H^{Non-Hecke,d}&=& \sum_{i-1}^{L-1} g_i - e^{i \gamma}\sin\gamma \left[ i - \tan\left(\f{\omega-\delta}{2} \right) \right] g^{d~(k_1)}_0 \nonumber \\
&& \quad -e^{i \gamma}\sin\gamma \left[ i + \tan\left(\f{\omega+\delta}{2} \right) \right] g^{d~(k_2)}_L 
\eea
where the non-Hecke boundary term $g^{(k_1,k_2)}_0$ was given in section \ref{se:NonDiagonalNonHeckeboundaries} and the diagonal terms $g^{d ~(k_1)}_0$ and $g^{d ~(k_2)}_L$ were given in \ref{se:Diagonalboundaries}. We recall from section \ref{se:NonDiagonalNonHeckeboundaries} that we always have $k_1+k_2 \le N$. In the extremal case in which $k_2=N-k_1$ the boundary generator $g_0$ reduces to a Hecke-type boundary and (\ref{eqn:NonHeckespectralequiv}) becomes the previous Hamiltonian (\ref{eqn:Heckespectralequiv}).

So far we have found spectral equivalences relating all one boundary chains to diagonal ones. However the space of diagonal chains that are
involved is limited to the ones satisfying $k_1 + k_2 \le N$. One might
wonder: what happens to the diagonal chains with $k_1 + k_2 > N$? As an
example consider the following $SU_q(3)$ diagonal boundary terms:
\bea \label{eqn:SU3example}
K_0^{d~(2)}=\left( \begin{array}{ccc}
1 & 0 & 0 \\
0 & 1 & 0 \\
0 & 0 & 0 
\end{array} 
\right)
\quad \quad 
K_L^{d~(2)}=\left( \begin{array}{ccc}
0 & 0 & 0 \\
0 & 1 & 0 \\
0 & 0 & 1 
\end{array} 
\right) 
\eea
These do not obey the constraint (\ref{eqn:OnebraidQuotient}). However let us consider the action of a parity operation, $P$, on the integrable chain. The bulk generators given in (\ref{eqn:SUNRmatrix}) transform as:
\bea
P g_i(q) = (1-q^2){\bf 1} +q^2 g_{L-i}(q^{-1})
\eea
where by $g_{L-i}(q^{-1})$ we mean $g_{L-i}$ with $q$ replaced everywhere by $q^{-1}$. One can easily verify algebraically that this is an automorphism of the Hecke algebra (\ref{eqn:Hecke}).  The boundary terms become:
\bea
P g_0^{d~(k_1)} = {\bf 1} - g_L^{d~(N-k_1)} \\
P g_L^{d~(k_2)} = {\bf 1} - g_0^{d~(N-k_2)}
\eea
In the $SU_q(3)$ example (\ref{eqn:SU3example}) we have:
\bea
&&P  \left( \begin{array}{ccc}
1 & 0 & 0 \\
0 & 1 & 0 \\
0 & 0 & 0 
\end{array}
\right) \otimes {\bf 1} \otimes {\bf 1} \cdots \otimes {\bf 1} = {\bf 1} \otimes {\bf 1} \cdots \otimes \left( \begin{array}{ccc}
1 & 0 & 0 \\
0 & 1 & 0 \\
0 & 0 & 0 
\end{array}
\right)= {\bf 1} - g_L^{d~(1)} \\
&&P {\bf 1} \otimes {\bf 1} \cdots \otimes \left( \begin{array}{ccc}
1 & 0 & 0 \\
0 & 1 & 0 \\
0 & 0 & 0 
\end{array}
\right)  = \left( \begin{array}{ccc}
1 & 0 & 0 \\
0 & 1 & 0 \\
0 & 0 & 0 
\end{array}
\right) \otimes {\bf 1} \cdots \otimes {\bf 1}  = {\bf 1} - g_0^{d~(1)} 
\eea
If the original boundary terms $g_0^{d~(k_1)}$ and $g_L^{d~(k_2)}$ have $k_1 + k_2 > N$ then it is easy to see that the terms $g_0^{d~(N-k_2)}$ and $g_L^{d~(N-k_1)}$ must satisfy $k_1 + k_2 \le N$. Therefore, up to constant terms, we have transformed our system to one obeying the constraint (\ref{eqn:OnebraidQuotient}). Now all the previous arguments can be used and we again have a spectral equivalence between the diagonal and one-boundary systems. Therefore all possible integrable diagonal boundary terms have spectral equivalences to one boundary chains.

In the next section we shall use the spectral equivalences to understand the patterns of $SU_q(N)$ symmetry breaking that can occur with the addition of an integrable boundary term.
\section{$SU_q(N)$ symmetry breaking}
\label{se:SUNbreaking}
The bulk $R$-matrix (\ref{eqn:SUNRmatrix}) is well known to be invariant under an $SU_q(N)$ quantum group symmetry\cite{Jimbo:1989qm}. As a consequence the bulk integrable theory also possesses this symmetry. In this section we shall discuss the effect of adding to this a single integrable boundary term.

In the first subsection we shall discuss the bulk symmetry. The different
possible integrable boundary terms that can be added to one end have been
discussed in sections \ref{se:NonDiagonalHecke} and
\ref{se:NonDiagonalNonHeckeboundaries}. However it is not immediately obvious
what symmetries they preserve. We shall use the spectral equivalences of
section \ref{sec:SpectralEquivs} to discuss this problem from the alternative
perspective of the diagonal chain where it is much easier to see the
symmetries. This reveals a pattern of symmetry breaking patterns by integrable
boundary conditions. We shall give explicit examples for $SU(2)$ and
$SU(3)$.

Throughout this section we shall work at generic values of the parameters. It is very likely that, as for the XXZ case \cite{Nichols:2004fb}, there can be enhancement of the symmetry for exceptional points but we shall not discuss this here.
\subsection{Quantum group symmetry}
The $SU_q(N)$ algebra is generated by $E^{\alpha}$, $F^{\alpha}$ and $q^{\pm H^{\alpha}/2}$ with $\alpha=1,\cdots,N-1$ subject to the relations:
\bea
q^{H^\alpha /2} E^{\beta} q^{-H^\alpha /2} &=& q^{a_{\alpha \beta}/2} E^{\beta} \nonumber \\
q^{H^\alpha /2} F^{\beta} q^{-H^\alpha /2} &=& q^{-a_{\alpha \beta}/2} E^{\beta} \nonumber\\
\left[E^{\alpha},F^{\beta} \right] &=& \delta_{\alpha \beta} \f{q^{H^{\alpha}}-q^{-H^{\alpha}}}{q-q^{-1}}
\eea
\bea
\left[E^{\alpha},E^{\beta} \right] = 0   \quad \left[F^{\alpha},F^{\beta} \right] = 0 \quad \quad {\rm if~} a_{\alpha \beta}=0 \nonumber
\eea
\bea
E^{\alpha} E^{\alpha} E^{\beta} -(q+q^{-1}) E^{\alpha} E^{\beta} E^{\alpha} + E^{\beta} E^{\alpha} E^{\alpha} &=&0 \quad \quad {\rm if~} a_{\alpha \beta}=-1 \nonumber\\
F^{\alpha} F^{\alpha} F^{\beta} -(q+q^{-1}) F^{\alpha} F^{\beta} F^{\alpha} + F^{\beta} F^{\alpha} F^{\alpha} &=&0 \quad \quad {\rm if~} a_{\alpha \beta}=-1 \nonumber
\eea 
The numbers $a_{\alpha \beta}$ are the entries of the Cartan matrix $A$. The co-products are given by:
\bea \label{eqn:SUNcoproducts}
\Delta(q^{\pm H^{\alpha}/2})&=&q^{\pm H^{\alpha}/2} \otimes q^{\pm H^{\alpha}/2} \nonumber\\
\Delta(E^{\alpha})&=&q^{H^{\alpha}/2} \otimes E^{\alpha} + E^{\alpha} \otimes q^{-H^{\alpha}/2} \\
\Delta(F^{\alpha})&=&q^{H^{\alpha}/2} \otimes F^{\alpha} + F^{\alpha} \otimes q^{-H^{\alpha}/2}\nonumber
\eea
Using these we have the following representations on the spin chain:
\bea
q^{\pm H^{\alpha}/2}&=&q^{\pm h^{\alpha}/2} \otimes \cdots \otimes q^{\pm h^{\alpha}/2} \nonumber\\
E^{\alpha} &=& \sum_{i} q^{h^{\alpha}/2} \otimes \cdots \otimes q^{h^{\alpha}/2} \otimes e^{\alpha}_i \otimes q^{-h^{\alpha}/2} \otimes \cdots \otimes q^{-h^{\alpha}/2} \\
F^{\alpha} &=& \sum_{i} q^{h^{\alpha}/2} \otimes \cdots \otimes q^{h^{\alpha}/2} \otimes f^{\alpha}_i \otimes q^{-h^{\alpha}/2} \otimes \cdots \otimes q^{-h^{\alpha}/2} \nonumber
\eea
where $e^{\alpha}$, $f^{\alpha}$, and $h^{\alpha}$ are the generators in the Chevalley basis of (non-quantum) $SU(N)$. We give below explicit forms in the case of $SU(2)$ and $SU(3)$.
\begin{itemize}
\item{$SU(2)$}

The Cartan matrix is just a number:
\bea
A=2
\eea
There is only one simple root and the generators of $SU(2)$ are given by:
\bea
h=\left( \begin{array}{cc}
1 & 0 \\
0 & -1
\end{array}
\right) \quad e=\left( \begin{array}{cc}
0 & 1 \\
0 & 0
\end{array}
\right) \quad
f=\left( \begin{array}{cc}
0 & 0 \\
1 & 0
\end{array}
\right)
\eea
\item{$SU(3)$}
The Cartan matrix is:
\bea
\left( \begin{array}{cc}
2 & -1 \\
-1 & 2 
\end{array} \right)
\eea
There are now two simple roots and the generators of $SU(3)$ are given by:
\bea
\label{eqn:SU3groupgens}
h^{1}=\left( \begin{array}{ccc}
1 & 0 & 0 \\
0 & -1 & 0 \\
0 & 0 & 0
\end{array}
\right) \quad e^{1}=\left( \begin{array}{ccc}
0 & 1 & 0 \\
0 & 0 & 0 \\
0 & 0 & 0
\end{array}
\right) \quad
f^{1}=\left( \begin{array}{ccc}
0 & 0 & 0 \\
1 & 0 & 0 \\
0 & 0 & 0
\end{array}
\right) \\
h^{2}=\left( \begin{array}{ccc}
0 & 0 & 0 \\
0 & 1 & 0 \\
0 & 0 & -1
\end{array}
\right) \quad 
e^{2}=\left( \begin{array}{ccc}
0 & 0 & 0 \\
0 & 0 & 1 \\
0 & 0 & 0
\end{array}
\right) \quad
f^{2}=\left( \begin{array}{ccc}
0 & 0 & 0 \\
0 & 0 & 0 \\
0 & 1 & 0
\end{array}
\right)
\eea
\end{itemize}

\subsection{Symmetry breaking by integrable boundary terms}
The presence of an integrable boundary term will destroy many of the bulk symmetries. We shall only examine the case of a single (left) boundary term. Using the spectral equivalences of the previous section we know we can discuss this problem in the context of the diagonal chain:
\bea
H=-a_0 g^{d~(k_1)}_0 -\sum_{i=1}^n g_i -a_L g^{d~(k_2)}_L
\eea 
where the $g^{d~(k_1)}_0$ and $g^{d~(k_2)}_L$ were defined in section
\ref{se:Diagonalboundaries}. It is clear, as both boundary terms are diagonal,
this Hamiltonian conserves all diagonal charges. However there may also be some additional non-Abelian conserved charges.

In the $SU(2)$ case there is only one integrable boundary term that can be added to the left end. It is given by (\ref{eqn:SU2Heckeboundary}):
\bea
K_0^{(1)}=\left( \begin{array}{cc}
1+r^2 & -ir \\
-ir & 0
\end{array}
\right)
\eea
The spectral equivalence of section \ref{sec:SpectralEquivs} relates this to a
chain with the diagonal boundary terms:
\bea
K_0^{d~(1)}=\left( \begin{array}{cc}
1 & 0 \\
0 & 0
\end{array}
\right)  
\quad \quad
K_L^{d~(1)}=\left( \begin{array}{cc}
0 & 0 \\
0 & 1
\end{array}
\right)
\eea
The only generators from the quantum group which commutes with these two
diagonal terms are $q^H$ and $q^{-H}$. Therefore $SU_q(2)$ is broken to a
single $U(1)$ charge. This is in agreement with the explicit
diagonalization\cite{Nichols:2004fb} of the boundary quantum group charge
\cite{Delius:2001qh,Delius:2002mv,Doikou:2004km}. We stress that throughout
this section we are working at generic values of the parameters where the
charges are always fully diagonalizable.

In the $SU(3)$ case there are three distinct integrable boundary terms. 
\begin{itemize}
\item

The first boundary term (\ref{eqn:SU31rr}) is given by:
\bea 
K_0^{(1)}=\left( \begin{array}{ccc}
1+r^2 & 0 & -ir \\
0 & r^2 & 0 \\
-ir & 0 & 0 
\end{array} 
\right)
\eea
In the isospectral diagonal chain this corresponds to the boundary terms:
\bea 
K_0^{d~(1)}=\left( \begin{array}{ccc}
1 & 0 & 0 \\
0 & 0 & 0 \\
0 & 0 & 0 
\end{array} 
\right) 
\quad \quad
K_L^{d~(2)}=\left( \begin{array}{ccc}
0 & 0 & 0 \\
0 & 1 & 0 \\
0 & 0 & 1 
\end{array} 
\right) 
\quad
\eea
It is obvious that all the diagonal generators commute with these.  However in this case there are additional non-Abelian symmetries. 

The generator $e^2$ of the $SU(3)$ group (\ref{eqn:SU3groupgens}) commutes
with both of these boundary matrices. Now using the co-products
(\ref{eqn:SUNcoproducts}), and the fact that all the diagonal charges are
conserved, we see that the generator $E^2$ commutes with both these boundary
generators. A similar argument applies to the generator $F^2$. Therefore the
$SU_q(3)$ quantum group is broken to $SU_q(2) \otimes U(1)$. In the next
subsection we shall present numerical results which confirm this pattern of
symmetry breaking.  
\item

The second boundary term corresponds to:
\bea 
K_0^{(2)}=\left( \begin{array}{ccc}
1+r^2 & 0 & -ir \\
0 & 1 & 0 \\
-ir & 0 & 0 
\end{array} 
\right) 
\eea
This is isospectral to a chain with diagonal boundary terms:
\bea 
K_0^{d~(1)}=\left( \begin{array}{ccc}
1 & 0 & 0 \\
0 & 0 & 0 \\
0 & 0 & 0 
\end{array} 
\right) 
\quad \quad
K_L^{d~(1)}=\left( \begin{array}{ccc}
0 & 0 & 0 \\
0 & 1 & 0 \\
0 & 0 & 1 
\end{array} 
\right) 
\quad
\eea
This is very similar to the previous one but now, in addition to the diagonal generators, it is $E^1$ and $F^1$ that also commute with it. Therefore the $SU_q(3)$ quantum group is again broken to $SU_q(2) \otimes U(1)$. 
\item
The third choice of boundary terms is:
\bea 
K_0^{(1,1)}=\left( \begin{array}{ccc}
1+r^2 & -ir  & 0 \\
-ir & 0 & 0 \\
0 & 0 & 0 
\end{array} 
\right)
\eea
This is isospectral to a chain with diagonal boundary terms:
\bea 
K_0^{d~(1)}=\left( \begin{array}{ccc}
1 & 0 & 0 \\
0 & 0 & 0 \\
0 & 0 & 0 
\end{array} 
\right) 
\quad \quad
K_L^{d~(1)}=\left( \begin{array}{ccc}
0 & 0 & 0 \\
0 & 0 & 0 \\
0 & 0 & 1 
\end{array} 
\right) 
\quad
\eea
There are now only the two diagonal generators conserved. Therefore the quantum group is broken to $U(1) \otimes U(1)$.
\end{itemize}
A similar pattern of symmetry breaking occurs with all the possible integrable boundary terms. For the Hecke boundary terms of section \ref{se:NonDiagonalHecke} with $K_0^{(k)}$ the symmetry breaking is:
\bea
SU_q(N) \rightarrow SU_q(N-k) \otimes SU_q(k) \otimes U(1)
\eea
This symmetry breaking in the diagonal chain was previously described in \cite{Doikou:1998ek}. For the non-Hecke boundary conditions of section \ref{se:NonDiagonalNonHeckeboundaries} with $K_0^{(k_1,k_2)}$ the symmetry breaking is:
\bea
SU_q(N) \rightarrow SU_q(k_1) \otimes SU_q(k_2) \otimes SU_q(N-k_1-k_2) \otimes U(1)^2
\eea
In both cases the extra $U(1)$ factors are necessary to ensure that we have the full set of diagonal generators.

The case of $SU(2)$, corresponding to the XXZ model, is almost trivial as the symmetry can only break to $U(1)$.
\subsection{Numerical example: $SU_q(3)$ symmetry breaking}
\label{se:Symbreaknumerics}
In this subsection we shall consider a numerical example confirming the
previous analysis of symmetry breaking. We consider integrable boundary terms added to the $SU_q(3)$ quantum group invariant Hamiltonian (\ref{eqn:BulkIntegrable}) on a chain of length $L=3$. 
In the first column of the table on page \pageref{tab:SUq3} the eigenvalues of the quantum group invariant Hamiltonian with $\gamma=0.3423$ ~($q=e^{i \gamma}$) are given. The degeneracies are as expected from the $SU(3)$ tensor product:
\bea
{\bf 3} \otimes {\bf 3} \otimes {\bf 3} = {\bf 1} \oplus {\bf 8} \oplus {\bf 8} \oplus {\bf 10}
\eea
The second column of the table gives the eigenvalues of the Hamiltonian
(\ref{eqn:Heckespectralequiv}) with the $g^{(1)}_0$ Hecke boundary term
(\ref{eqn:SU31rr}) and $\omega=0.54564$, $\delta=-i$. One can see that the quantum group eigenvalues split in the following way:
\bea
{\bf 1} &\rightarrow&  {\bf 1} \nonumber\\
{\bf 8} &\rightarrow&  {\bf 1} \oplus {\bf 2} \oplus {\bf 2} \oplus {\bf 3} \\
{\bf 10} &\rightarrow&  {\bf 1} \oplus {\bf 2} \oplus {\bf 3} \oplus {\bf 4} \nonumber
\eea
These are exactly the branching rules expected from breaking the $SU(3)$
symmetry to an $SU(2)$ subgroup. This is confirmed by repeating the exercise
with different lengths of chain. In particular the fundamental representation and
its conjugate give:
\bea
{\bf 3} &\rightarrow&  {\bf 1} \oplus {\bf 2} \\
{\bf \bar{3}} &\rightarrow&  {\bf 1} \oplus {\bf 2} \nonumber
\eea
The presence of extra Abelian factors in the broken symmetry is
not possible to detect by simply observing the spectrum as they do not give rise to
degeneracies. 

In the third column of the table we have given eigenvalues of the Hamiltonian
(\ref{eqn:NonHeckespectralequiv}) for the case of the $g^{(1,1)}_0$ non-Hecke
boundary term (\ref{eqn:NonHeckeSU311}) with again $\omega=0.54564$, $\delta=-i$. Now one can see that
there are no degeneracies consistent with the fact that the quantum group is
broken to purely Abelian factors.

\begin{table}[ht]
\label{tab:SUq3}
\begin{center}
\begin{tabular}{c|c|c}
 & \multicolumn{2}{c}{Integrable boundary term} \\
Quantum Group &  Hecke& Non-Hecke\\
\hline
-1.5493 - 1.2647 i & -1.5728 - 1.4548 i & -1.5752 - 1.3903 i \\
-0.7167 - 0.9680 i & -0.7277 - 1.1536 i & -0.7345 - 1.1230 i \\
-0.7167 - 0.9680 i & -0.7277 - 1.1536 i & -0.7276 - 0.9993 i \\
-0.7167 - 0.9680 i & -0.7189 - 1.1505 i & -0.7189 - 1.1505 i \\
-0.7167 - 0.9680 i & -0.7189 - 1.1505 i & -0.7026 - 0.9961 i \\
-0.7167 - 0.9680 i & -0.7189 - 1.1505 i & -0.6767 - 1.0531 i \\
-0.7167 - 0.9680 i & -0.6283 - 1.1182 i&  -0.6463 - 1.0907 i \\
-0.7167 - 0.9680 i & -0.6283 - 1.1182 i & -0.6283 - 1.1182 i \\
-0.7167 - 0.9680 i & -0.6159 - 1.1138 i  &-0.6190 - 1.0876 i\\
 ~1.1673 - 0.2967 i &  ~1.1563  - 0.4823 i&  ~1.1315 - 0.3870 i\\
 ~1.1673 - 0.2967 i &  ~1.1563  - 0.4823 i&  ~1.1331  - 0.3931 i\\
 ~1.1673 - 0.2967 i &  ~1.1814  - 0.4734 i&  ~1.1814  - 0.4734 i \\
 ~1.1673 - 0.2967 i &  ~1.1814  - 0.4734 i&  ~1.1833  - 0.4727 i \\
 ~1.1673 - 0.2967 i &  ~1.1833  - 0.4727 i & ~1.1902  - 0.3879 i \\
 ~1.1673  - 0.2967 i & ~1.1833  - 0.4727 i & ~1.1904  - 0.4362 i \\
 ~1.1673  - 0.2967 i & ~1.1833  - 0.4727 i & ~1.2057  - 0.3805 i \\
 ~1.1673  - 0.2967 i & ~1.1939  - 0.4689 i & ~1.2084  - 0.3863 i\\
 ~2.0000 + 0.0000 i &    ~1.9890  - 0.1856 i&  ~1.9890  - 0.1856 i \\
 ~2.0000 + 0.0000 i &    ~1.9890  - 0.1856 i & ~2.0000  + 0.0000 i  \\
 ~2.0000 + 0.0000 i &    ~1.9890  - 0.1856 i & ~2.0317  - 0.1260 i \\
 ~2.0000 + 0.0000 i &    ~1.9890  - 0.1856 i & ~2.0342  - 0.0579 i \\
 ~2.0000 + 0.0000 i &    ~2.0901  - 0.1496 i & ~2.0707  - 0.0192 i \\
 ~2.0000 + 0.0000 i &    ~2.0901  - 0.1496 i & ~2.0901  - 0.1496 i \\
 ~2.0000 + 0.0000 i &    ~2.0901  - 0.1496 i&  ~2.0994  - 0.0809 i \\
 ~2.0000 + 0.0000 i &    ~2.1259  - 0.1368 i&  ~2.1157  - 0.0704 i\\
 ~2.0000 + 0.0000 i &   ~2.1383  - 0.1324 i & ~2.1259  - 0.1368 i \\
 ~2.0000 + 0.0000 i &   ~2.1383  - 0.1324 i&  ~2.1383  - 0.1324 i\\ 
\end{tabular}
\end{center}
\caption{Eigenvalues of an $SU_q(3)$ Hamiltonian at $L=3$ sites with Hecke and
  non-Hecke integrable boundary terms. The values of parameters used are given
  in subsection \ref{se:Symbreaknumerics}.}
\end{table}
\section{Conclusion}
In this paper we have considered the $SU_q(N)$ model with different types of integrable boundary terms. The single most important result is that the $SU_q(N)$ model with any integrable non-diagonal boundary term added to one end is iso-spectral to the same $SU_q(N)$ model with purely diagonal boundary terms added to both ends and vice-versa.

In section \ref{se:1BTLreview} we reviewed the spectral equivalence found between the XXZ model with diagonal terms on both ends and the same XXZ model with an arbitrary left boundary term \cite{Nichols:2004fb}. The results at a low number of sites (\ref{eqn:diagonal1BTL}) motivated the study of the canonical diagonal representation. To understand this structure we required the braid group and Hecke algebras which were given in section \ref{se:Braid}. In section \ref{se:SUN} we introduced the integrable $SU_q(N)$ model and its possible integrable boundary terms. The algebraic description gives one a very transparent way to understand the structure of possible integrable boundary terms. In section \ref{se:Diagonal} we gave the general structure of the canonical diagonal representation. The existence of this suggested spectral equivalences between $SU_q(N)$ models with a single integrable non-diagonal boundary term and those with two diagonal boundaries. Although we have not proved the equivalences of section \ref{sec:SpectralEquivs} they are consistent with a large number of numerical checks. In section \ref{se:SUNbreaking} we have used these spectral equivalences to discuss the structure of symmetry breaking that can occur when an integrable boundary term is added to the  $SU_q(N)$ model.

There are many outstanding questions raised by this paper. The most urgent is
undoubtedly a proof of the spectral equivalences given in section
\ref{sec:SpectralEquivs}. In this paper the structure of the real diagonal
representation, an example of which is given in
(\ref{eqn:1BTLDiagonalLocalizedRepn}), was not discussed. This appears to be a
less algebraic object than the canonical diagonal representation as it also
involves the parameters of the Hamiltonian. 

The symmetry breaking patterns
presented in section \ref{se:SUNbreaking} are for the case of generic
parameters only. The structure of
conserved charges in one-boundary systems has been discussed recently in the
literature \cite{Delius:2001qh,Delius:2002mv,Doikou:2004km} generalizing
earlier results at free fermion point \cite{Mezincescu:1997nw}. One would expect, as with the XXZ case discussed in
\cite{Nichols:2004fb}, that, although the spectral equivalences continue to
hold at all points, the diagonal and one-boundary Hamiltonians can have
different indecomposable structure. This will occur at the points which the
relevant algebra becomes non-semisimple. For the Hecke algebra of type B
(\ref{eqn:BtypeHecke}) this is known to be when $r$ is a power of $q$. For the
cubic quotient, defined by (\ref{eqn:cubic}) and (\ref{eqn:extraquotient}), we do not know the corresponding points but it is natural to
conjecture that they might be given by the same set. Throughout this paper
when we have considered integrable boundaries on both ends we have restricted
our consideration to the very special case of two diagonal boundaries. In some cases, but not the XXZ one, one expects part of
the quantum group symmetry to survive even for non-diagonal boundary terms at
both ends. Finally, it would be extremely interesting to understand extensions of our results to other integrable models.

\renewcommand\thesection{}
\section{Acknowledgements}
I am grateful for financial support by the E.U. network {\it Integrable models and
  applications: from strings to condensed matter} HPRN-CT-2002-00325. I would like to
  thank V. Rittenberg for many stimulating discussions. I would also like to
  thank P. Pyatov and A. Ram for discussions on braid and Hecke algebras. 
%
%
\appendix
\renewcommand{\theequation}{\Alph{section}.\arabic{equation}}

\setcounter{equation}{0} 
\section{Link pattern representation of 1BTL}
\label{se:Loop}

A different representation of the 1BTL algebra, from the spin chain, in another $2^L$ dimensional
vector space is obtained if we use link patterns\cite{JanReview}.

We start by giving the standard graphical representation of the 1 BTL algebra:
\bea
e_i \quad &=&\quad
\begin{picture}(130,20)
\put(0,0){\epsfxsize=130pt\epsfbox{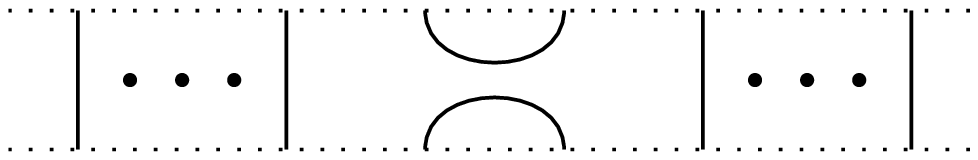}}
\put(51,-10){{\small i}}
\put(67,-10){{\small i+1}}
\end{picture} \\
\nonumber \\
e_0 \quad &=&\quad
\begin{picture}(240,20)
\put(0,0){\epsfxsize=70pt\epsfbox{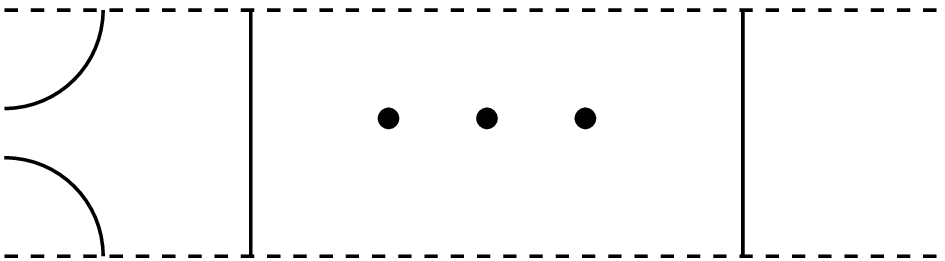}}
\put(40,-10){{\small 1}}
\end{picture}
\label{eq:f1}
\eea
Multiplication of two words in the algebra corresponds to putting one
word below the other and merging the loops lines. For example, the
relations $e_i^2=2 \cos \gamma e_i$, $e_0^2=\f{\sin \omega}{\sin (\omega+\gamma)} e_0$ and $e_ie_{i+1}e_i=e_i$ graphically read:
\bea
\begin{picture}(120,40)
\put(0,0){\epsfxsize=40pt\epsfbox{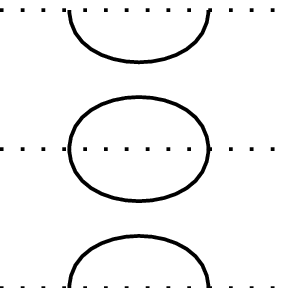}}
\put(50,18){$=2 \cos \gamma$}
\put(100,10){\epsfxsize=40pt\epsfbox{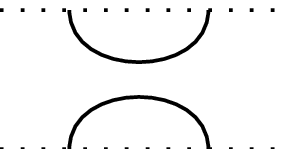}}
\end{picture}
\label{eq:e^2}
\eea
\vskip2mm
\bea
\begin{picture}(240,20)
\put(0,0){\epsfxsize=70pt\epsfbox{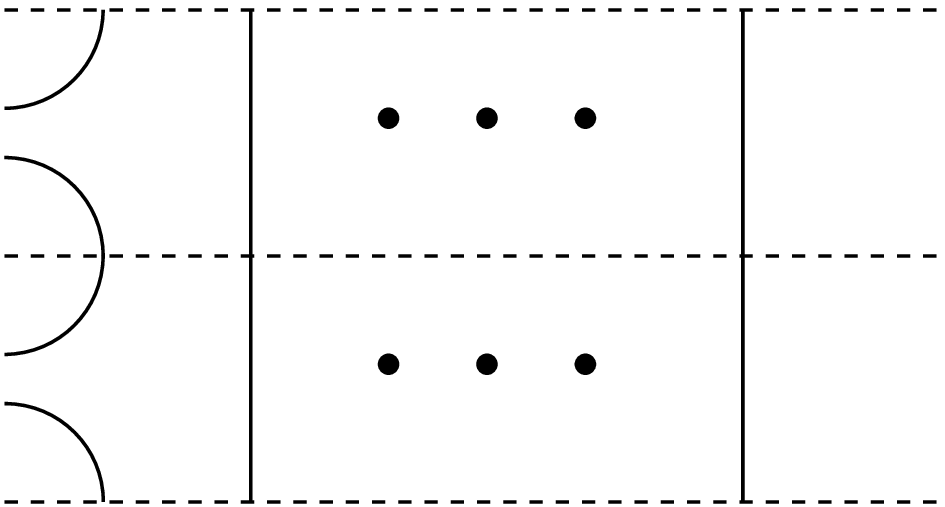}}
\put(80,18){$=\f{\sin \omega}{\sin(\omega+\gamma)}$} 
\put(140,10){\epsfxsize=70pt\epsfbox{e0.eps}}
\end{picture}
\label{eqn:e0^2}
\eea
\vskip2mm
\bea
\begin{picture}(150,60)
\put(0,0){\epsfxsize=60pt\epsfbox{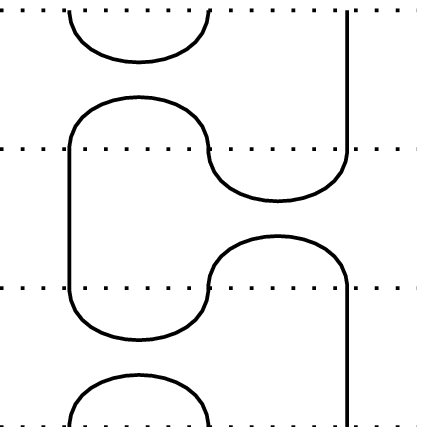}}
\put(70,28){$=$}
\put(90,20){\epsfxsize=60pt\epsfbox{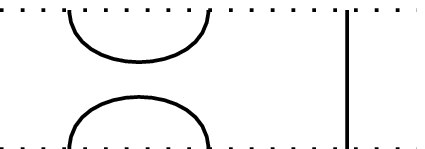}}
\end{picture}
\eea
\vskip 6mm
The link pattern representation corresponds to considering an ideal of
the 1BTL \cite{JanReview}. We consider the state $\ket$
 by taking the graph corresponding to the unit element $\bf{1}$ of the 1BTL algebra. It
 has all $L$ sites unconnected:
\vskip 2mm
\bea
\ket  \quad = \quad \begin{picture}(140,10)
\put(0,0){\epsfxsize=70pt\epsfbox{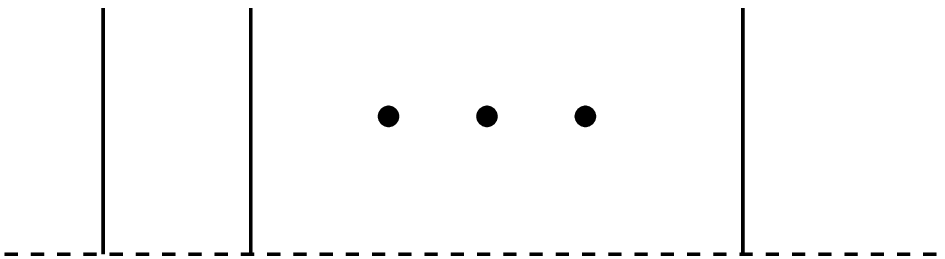}}
\end{picture}
\eea
We then act with the algebra and finally keep only the bottom half of the picture:

\bea
{\bf 1} \ket  \quad = \quad \begin{picture}(140,10)
\put(0,0){\epsfxsize=70pt\epsfbox{Ideal.eps}}
\end{picture}
\eea
\bea
e_i \ket \quad = \quad
\begin{picture}(140,10)
\put(0,0){\epsfxsize=140pt\epsfbox{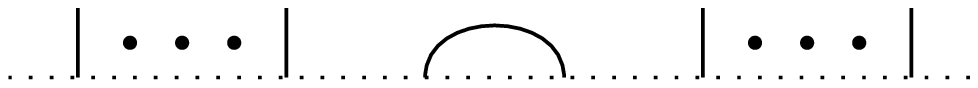}}
\put(56,-10){{\small i}}
\put(72,-10){{\small i+1}}
\end{picture}
\label{eq:eiT} \\
\nonumber \\
e_0 \ket \quad= \quad 
\begin{picture}(140,10)
\put(0,0){\epsfxsize=70pt\epsfbox{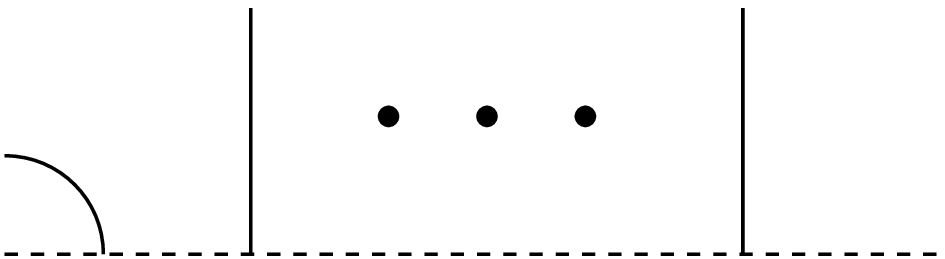}}
\end{picture}
\eea
\vskip 6mm
Other examples are:
\bea
e_{i+1}e_i \ket \quad &=&\quad 
\begin{picture}(160,30)
\put(0,0){\epsfxsize=160pt\epsfbox{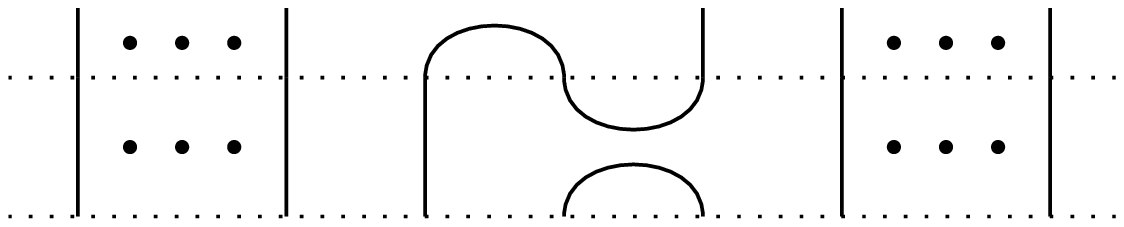}}
\put(56,-10){{\small i}}
\put(72,-10){{\small i+1}}
\put(92,-10){{\small i+2}}
\end{picture} \nonumber\\[14pt]
\nonumber \\
&=& \quad
\begin{picture}(160,10)
\put(0,0){\epsfxsize=160pt\epsfbox{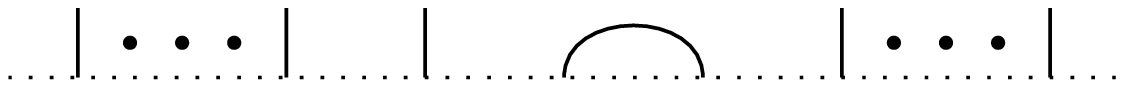}}
\put(56,-10){{\small i}}
\put(72,-10){{\small i+1}}
\put(95,-10){{\small i+2}}
\end{picture}
\label{eq:eip1T}
\eea
\vskip 6mm
and:
\bea
e_{i+1}e_{i+2}e_i \ket \quad = \quad
\begin{picture}(170,15)
\put(0,0){\epsfxsize=170pt\epsfbox{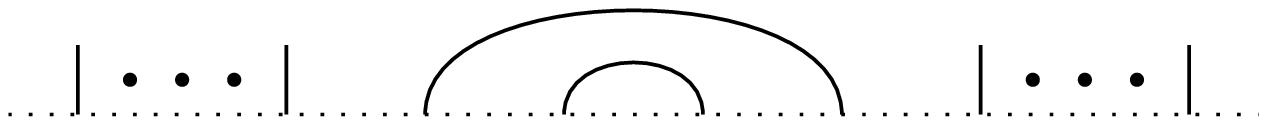}}
\put(52,-10){{\small i}}
\put(106,-10){{\small i+3}}
\end{picture}\;.
\label{eq:eiTcomp}
\eea
\vskip 6mm
It is natural to introduce a diagrammatic charge, $C$, which counts the number of
sites which contain loops connected to another site or to the boundary. This
diagrammatic charge has \emph{no} algebraic
properties and so does not see any particular structure in the 1BTL algebra at
the critical points. In general the link patterns form a vector space of dimension $2^L$ and the
diagrammatic charge $C$ splits it into subspaces having the dimension of the
binomial coefficients. More specifically for a system of size $L$ we have
$C=0,1,\cdots,L$ with:
\begin{itemize}
\item{$C$ even}
\bea
{\rm dimension}=\left(
\begin{array}{c}
L \\
\f{L}{2} - C 
\end{array}
\right)
\eea
\item{$C$ odd}
\bea
{\rm dimension}=\left(
\begin{array}{c}
L \\
-\f{L}{2} + C - 1  
\end{array}
\right)
\eea
\end{itemize}
However it is important to note that the diagrammatic charge $C$ does
\emph{not} commute with the 1BTL generators. However we can define a
\emph{new} representation in which it does. First note that, as all elements of the algebra cannot break links (i.e. decrease $C$), we
have a natural decomposition when acting on states of charge $c$:
\bea
e_i \state{c} = \tilde{e_i} \state{c} + \left\{{\rm States~with~ C>c}\right\}
\eea
It is easy to see that the $\tilde{e_i}$ so defined will obey the
1BTL \emph{and} will commute with $C$. We call this the chopped diagonal representation.

At a low number of sites we found that this chopped diagonal representation is equivalent at all points to the 1BTL canonical diagonal representation arising in the spin chain.
\setcounter{equation}{0} 
\section{Baxterization}
\label{se:Baxterization}
In this section we discuss the process of Baxterization \cite{Jones:1990hq}
for the different boundary conditions that we have given in this paper. The
idea is that if one has a solution to the Yang-Baxter or reflection equations
in the braid limit (i.e. without spectral parameters) then one may try to
reconstruct the full solution with spectral parameters. For the purposes of this paper the form of the full
solutions is only required to derive the corresponding integrable chains. For
the Yang-Baxter \cite{Jones:1990hq} and Hecke type boundary \cite{Doikou:2002ry} the solutions have previously appeared in the literature and we include them only for completeness. The non-Hecke Baxterization is, to our knowledge, new.
\subsection{Yang-Baxter equation}
\label{se:FullYB}
If the bulk generators $g_i$ obey the Hecke algebra (\ref{eqn:Hecke}) then using the ansatz:
\bea \label{eqn:FullYBsoln}
R_i(u)={\bf 1}-f(u)g_i
\eea
one finds that $R_i(u)$ obeys the Yang-Baxter equation if:
\bea
(g_2-g_1) \left(f(u)+f(v)+(q^2-1)f(u)f(v)-f(u+v) \right)=0
\eea
Let us illustrate how to solve this as all cases proceed in a similar manner. First by putting $u=0$ we find non-trivial solutions require $f(0)=0$. Now differentiating w.r.t $u$ and then putting $u=0$ we get:
\bea
f'(0)+(e^{2 i \gamma}-1)f(v) f'(0)-f'(v)=0
\eea
where $q=e^{i \gamma}$. This is a simple first order equation. The value of $f'(0)$ is simply a scale and with the choice $f'(0)=\f{2i}{e^{2 i \gamma}-1}$ we get the solution:
\bea      
f(u)=\f{e^{2 i u}-1}{e^{2 i \gamma}-1}
\eea
The resulting $R_i(u)$ also obeys the so-called unitarity condition:
\bea
R_i(u) R_i(-u) = \left(1-\f{\sin^2 u}{\sin^2 \gamma} \right) \bf{1}
\eea
The corresponding integrable chain is given by:
\bea
H&=&\sum_{i=1}^{L-1} R'_i(0) \nonumber \\
&=& \f{2i}{e^{2 i \gamma}-1} \sum_{i=1}^{L-1} g_i
\eea
\subsection{Reflection equation}
\subsubsection{Hecke type boundary}
\label{se:FullHeckeBYB}
In the case in which $g_0$ is a generators of the Hecke algebra of type B (\ref{eqn:BtypeHecke}) and $R_i(u)$ is given in section (\ref{se:FullYB}) we make the ansatz:
\bea \label{eqn:HeckeAnsatz}
K(u)={\bf 1}-a(u)g_0
\eea
one finds that $K(u)$ obeys the reflection equation if:
\bea
\Bigl\{e^{2 i \omega} f(u+v)\left(a(u)-a(v)\right)+f(u-v)\left(-e^{2 i \omega} (1+(-1+q^2)f(u+v)\right)a(v)\nonumber \\
+a(u)\left((1+e^{2i \omega})a(v)-e^{2i \omega}\right)\Bigr\}(g_1 g_0 - g_0 g_1 )=0
\eea
This gives:
\bea
a(u)=\f{e^{2 i \omega} \left(e^{4 i u}-1 \right)}{1+e^{2 i \omega} +e^{i(\omega+\delta+2u)}+e^{i(\omega-\delta+2u)}}
\eea
where the coefficient $\delta$ is arbitrary. The corresponding $K$-matrix (\ref{eqn:HeckeAnsatz}) also satisfies:
\bea
K(u) K(-u)= \f{2\left( \cos \delta + \cos(2u-\omega) \right) \left(\cos \delta + \cos(2u+\omega) \right)}{2+\cos 2 \delta + 4 \cos \delta \cos 2u \cos \omega + \cos 2 \omega } {\bf 1}
\eea
The integrable chain is given by:
\bea
H&=&\sum_{i=1}^{L-1} R'_i(0)+\f{1}{2} K'(0) \nonumber \\
&=& \f{2i}{e^{2 i \gamma}-1} \left\{ -\f{i e^{i(\gamma+\omega)} \sin \gamma }{\cos \omega + \cos \delta }~ g_0 + \sum_{i=1}^{L-1} g_i \right\}
\eea
\subsubsection{Non-Hecke type boundary}
\label{se:FullNonHeckeBYB}
In the case in which $g_0$ satisfies (\ref{eqn:BtypeHecke}) and $R_i(u)$ is given in section (\ref{se:FullYB}) we make the ansatz:
\bea \label{eqn:NonHeckeAnsatz}
K(u)&=&{\bf 1}-a(u)g_0 -b(u) g^2_0
\eea
Then after a great deal of algebra and using only the reflection equation and Hecke condition for $g_1$ we find that this leads to:

%
%
%
\bea \label{eqn:simplifyingeqn}
(g_1 g_0 -g_0 g_1) A + (g_1 g_0^2 -g_0^2 g_1) B + (g_0 g_1 g_0^2 -g_0^2 g_1 g_0 ) C \nonumber \\
+(g_1 g_0^2 g_1 g_0^2 - g_0^2 g_1 g_0^2 g_1)D=0
\eea
where:
\bea
A&=& f(u-v) a(u) - f(u+v) a(u) + f(u-v) a(v) + f(u+v) a(v)\nonumber \\
&&-f(u-v)f(u+v)a(v)(1-q^2)+r^2 f(u-v)a(v)b(u)\\
&&+ r^2 f(u-v)a(u)b(v)+r^2(1+r^2) f(u-v)b(u)b(v) \nonumber \\
B&=&-f(u-v)a(u)a(v)+f(u-v)b(u)-f(u+v)b(u)\nonumber \\
&&-(1+r^2)f(u-v)a(v)b(u)+f(u-v)b(v)+f(u+v)b(v)\nonumber \\
&&-(1-q^2)f(u-v)f(u+v)b(v)-(1+r^2)f(u-v)a(u)b(v) \\
&&-(1+r^2+r^4)f(u-v)b(u)b(v)\nonumber \\
C&=&  f(u+v) a(v)b(u) - (1-q^2)f(u+v)f(u-v)a(v)b(u) \nonumber \\
&&- f(u+v) a(u) b(v)  \\
D&=& f(u+v)f(u-v)b(u)b(v)
\eea
Clearly if all the terms in (\ref{eqn:simplifyingeqn}) were independent we
would only have the trivial solution $a(u)=b(u)=0$. However, as discussed in
section \ref{se:NonDiagonalNonHeckeboundaries} this is not the case as we have
an additional relation:
\bea \label{eqn:extraquotient1}
(g_1 g_0^2 g_1 g_0^2 -g_0^2 g_1 g_0^2 g_1)=(1+r^2)(g_1 g_0^2 g_1 g_0 -g_0 g_1 g_0^2 g_1)
\eea
With this extra relation, and using the Hecke condition for $g_1$ and reflection equation once again, we are able to combine the $C$ and $D$ terms:
\bea \label{eqn:thirdeqn}
&&(g_0 g_1 g_0^2 -g_0^2 g_1 g_0 ) C+(g_1 g_0^2 g_1 g_0^2 - g_0^2 g_1 g_0^2 g_1)D \nonumber\\
&&\quad=(g_0 g_1 g_0^2 -g_0^2 g_1 g_0 )\Bigr\{f(u+v) a(v)b(u) - (1-q^2)f(u+v)f(u-v)a(v)b(u) \nonumber \\
&&\quad \quad - f(u+v) a(u) b(v) +(1-q^2)(r^2+1) f(u+v)f(u-v)b(u)b(v)\Bigl\}
\eea
In order for (\ref{eqn:NonHeckeAnsatz}) to satisfy the full reflection equation this must vanish in addition to $A=B=0$. Using the solution for $f(u)$ and solving first $A=0$ and $B=0$ we get a solution for $a(u)$ and $b(u)$ with arbitrary coefficients $a_1=a'(0)$ and $b_1=b'(0)$. Now the vanishing of equation (\ref{eqn:thirdeqn}) requires an additional relation between $a_1$ and $b_1$:
\bea
\left( a_1+b_1(1+r^2) \right)\left(a_1^2+r^2 b_1^2 -4i b_1 +(r^2+1) a_1 b_1 \right)=0
\eea
The solution $a_1=-(1+r^2)b_1$ is not new as we have $(1+r^2)g_0-g_0^2=g_0^d$ and so it is just a diagonal solution. The other two quadratic solutions are new. They can be parameterized by:
\bea
a_1&=&\f{2i e^{i\omega}}{\cos \omega + \cos \delta}  \quad \quad 
b_1=\f{2i e^{-i(\pm \delta-2 \omega)}}{\cos \omega + \cos \delta} 
\eea
where $\delta$ is arbitrary. Clearly these are trivially related and we shall choose only the $(+)$ sign. This leads to the following solution for $a(u)$ and $b(u)$:
\bea
a(u)=\f{
e^{2i \omega}\left(e^{4iu}-1 \right) \left(e^{2iu}-1-e^{2i\omega}+e^{2i(\omega+u)}+e^{i(\delta+\omega+2u)}\right)
}
{\left(
e^{2iu}-1+e^{2i(u+\omega)}+ e^{i(\delta+\omega+2u)}
\right)
\left(
e^{2iu}-e^{2i\omega}+e^{2i(\omega+u)}+e^{i(\delta+\omega+2u)} \right)
}\\
b(u)=\f{
e^{4i \omega}\left(e^{4iu}-1 \right) 
}
{\left(
e^{2iu}-1+e^{2i(u+\omega)}+ e^{i(\delta+\omega+2u)}
\right)
\left(
e^{2iu}-e^{2i\omega}+e^{2i(\omega+u)}+e^{i(\delta+\omega+2u)} \right)
}
\eea
The corresponding $K$-matrix (\ref{eqn:NonHeckeAnsatz}) also satisfies:
\bea
K(u) K(-u)={\bf 1}
\eea
Finally we find the integrable chain is given by:
\bea
H&=&\sum_{i=1}^{L-1} R'_i(0)+\f{1}{2} K'(0) \nonumber \\
&=& \f{2i}{e^{2 i \gamma}-1} \Biggl\{ -i \f{e^{-i(\delta-\gamma)} \sin \gamma }{\cos \omega + \cos \delta} \left( 1+e^{i(\omega+\delta)} +e^{2i \omega} \right) g^{(k_1,k_2)}_0  \nonumber \\
&&\quad +i \f{e^{-i(\delta-\gamma)} e^{2i \omega} \sin \gamma }{\cos \omega + \cos \delta}  (g^{(k_1,k_2)}_0)^2  + \sum_{i-1}^{L-1} g_i \Biggr\}
\eea
After this derivation two facts should be emphasized. Firstly the additional quotient 
(\ref{eqn:extraquotient1}) was essential in order to carry out the Baxterization and obtain the full $K(u)$ matrix. Secondly, once $g_i$ and $g_0$ are fixed, the integrable Hamiltonian has only \emph{one} free parameter $\delta$.
%
%
%

\end{document}